\documentclass[12pt,preprint]{aastex}

\slugcomment{Not to appear in Nonlearned J., 45.}

\shorttitle{Non-thermal emissions from the Crab-like pulsars}
\shortauthors{Takata and Chang}

\begin{document}


\title{PULSE PROFILES, SPECTRA AND POLARIZATION CHARACTERISTICS
 OF NON-THERMAL EMISSIONS FROM THE CRAB-LIKE PULSARS}


\author{J.Takata}
\affil{Institute of Astronomy and Astrophysics, Academia Sinica; and 
 Theoretical Institute for Advanced Research in Astrophysics, 
Academia Sinica and National Tsing Hua University,  
Taipei Taiwan}
\and 
\author{H.-K.Chang}
\affil{Department of Physics and Institute of Astronomy, National 
Tsing Hua University, Hsinchu, Taiwan}



\begin{abstract}
We discuss non-thermal emission mechanism of the Crab-like pulsars 
with both a two-dimensional  electrodynamical study and a three-dimensional 
model. We investigate the emission process in the outer gap accelerator. 
In the two-dimensional electrodynamical study, we solve the Poisson equation
 of the accelerating electric field in the outer gap and 
 the equation of motion of the primary particles with 
the synchrotron and the curvature radiation process and
 the pair-creation process. We show a solved gap structure which produces 
a consistent gamma-ray spectrum with EGRET observation. Based on the 
two-dimensional model, we conduct a three-dimensional emission model 
to calculate the synchrotron and the inverse-Compton processes of the 
secondary pairs produced outside the outer gap. 
 We calculate  the pulse profiles, the phase-resolved spectra and 
the polarization characteristics in optical to $\gamma$-ray bands 
to compare the observation of the  Crab pulsar and PSR B0540-69. 
For the Crab pulsar, we find that the outer gap geometry extending 
from near the stellar surface to near the light cylinder produces 
a complex morphology change of the pulse profiles
 as a function of the photon energy. 
This predicted morphology change is quite similar with that of 
 the observations. 
The calculated phase-resolved spectra  are consistent 
with the data through optical to 
the $\gamma$-ray bands.
We demonstrate that the 10$\sim$20~\% of the polarization degree 
in the optical emissions from the Crab pulsar and the Vela pulsar are 
explained by the synchrotron emissions with the particle gyration motion. 
 For PSR B0540-69, 
the observed pulse profile with a single broad pulse is  reproduced with 
a thicker emission region and a smaller inclination angle  
between the rotational axis and the magnetic axis than the Crab pulsar. 
\end{abstract}




\section{Introduction}
The observed  strong  $\gamma$-ray emissions  from the 
seven young pulsars (Thompson 2003) 
show  that electrons and positrons 
are accelerated up to ultra-relativistic regime in the pulsar
magnetosphere.  The Crab pulsar (PSR B0531+21), 
which is  one of the  brightest and the youngest 
$\gamma$-ray emitting pulsar, shows  the non-thermal 
emission properties in optical to $\gamma$-ray bands. 
  The observed spectrum of the pulsed photons emitted from 
the Crab pulsar  extends continuously
 from optical to $\gamma$-ray bands with 
the spectral index $\alpha_{\nu}$, defined as 
$I_{\nu}\propto \nu^{-\alpha_{\nu}}$,  
varying from $\alpha_{\nu}\sim0$ in optical wavelengths, 
$\alpha_{\nu}\sim0.5$ 
in $X$-ray bands, to $\alpha_{\nu}\sim2$ in $\gamma$-ray bands. 
The pulse profile has two peaks in a single period,  
and the positions of the pulse peaks across the
wide energy range are approximately  all in phase (Kuiper et al. 2001). 
Interestingly, the pulse profile morphology 
changes significantly as a function of
the photon energy. The first peak (denoted Peak~1 in the following) 
dominates  in optical wavelengths. However, 
the second peak (Peak~2) becomes more and more pronounced for increasing 
energies and eventually the  Peak~2 emission dominates in soft
$\gamma$-ray bands.  Above 10~MeV photon energy,  
 Peak~1 again dominates  Peak~2. 
The electromagnetic spectrum of the non-thermal emissions  also 
 changes with pulse phases. 
 In the future,  the phase-resolved spectra
above 10~MeV will be measured with a sensitivity better  
than that of the  Energetic Gamma-Ray Experiment
Telescope  on board the \textit{Compton Gamma-ray Observatory} by, 
for example, GLAST LAT.  These observed detail properties 
 for the pulse profiles and the phase-resolved spectra will be useful 
to discriminate the proposed emission models.

In addition to the pulse profiles, 
 Kanbach et al.(2005) measured 
 the polarization characteristics of the pulsed photons from the Crab pulsar 
 in the optical wavelengths. The observation revealed that 
  the degree of the polarization at each  pulse
 peak is lower than 10~\% and a 
large swing of the position angle of the electric-vector of the radiation 
appears at each pulse peak. 
The polarization measurements provide two additional 
observed properties, namely,  
the degree  and the position angle of the polarization.  
In the future,  the polarization of  $X$-ray and soft
$\gamma$-ray emissions from the pulsar 
will probably be able to be measured by ongoing projects 
such as PoGO (Kataoka et al. 2005) and NCT (Chang et al. 2007) 
projects. Therefore,  
a theoretical study, particularly on the polarization characteristics, 
is not only desired, but also timely for the understanding
  the non-thermal emission process in the pulsar magnetospheres.

The polar cap accelerator 
(Ruderman and Sutherland 1975; Daugherty and Harding 1996) 
and the outer magnetospheric  accelerator, the   so called  outer gap model 
(Cheng et al. 1986a,b; Romani 1996),  were proposed  as 
 the possible  acceleration sites  in the pulsar magnetospheres. 
The traditional polar cap model assumes an acceleration region expanding 
 several stellar radii from the stellar surface around the magnetic pole. 
On the other hand, the traditional outer gap assumes an acceleration region 
extending beyond the null surface of the Goldreich-Julian charge 
density at the outer magnetosphere. 
The Goldreich-Julian charge density is given by  $\rho_{GJ}\sim
-\mathbf{\Omega}\cdot\mathbf{B}/2\pi c$ (Goldreich and Julian 1969)
with $\Omega$ being the rotational frequency of the star, 
$\mathbf{B}$  the magnetic field, and $c$ the speed of light. 
Both models assume the particle acceleration by  
an electric field parallel to the magnetic field line. In the pulsar
 magnetosphere, the accelerating electric field arises 
in the region where the local 
charge density differs from the Goldreich-Julian charge density. 

The slot gap model (Muslimov \& Harding 2004),
 which  is an extended polar cap model, 
predicts that the acceleration region extends up
 to near the light cylinder around the last-open 
field lines because the pair-formation front, 
which screens the accelerating electric field,
 occurs at higher altitude around there. 
 Two-dimensional electrodynamical studies 
 (Takata et al. 2004, 2006; Hirotani 2006) 
 suggested that the inner boundary of the outer  gap 
locates near (or at) the stellar surface because 
of the current through the outer gap. 
Although  the recent polar-slot gap  
and outer gap models both predict similar geometry of 
the acceleration region, 
an important difference between the two models is  the electric field 
configuration   
in the accelerator. For the slot gap accelerator, the electric 
field is stronger nearer the stellar surface and smaller at higher altitude. 
On the other hand, the outer gap model 
predicts a stronger electric 
field beyond the null surface and a smaller one below the null surface 
 due to the screening effect of  electron and positron pairs. 
This difference in the electric field configuration, and the resultant
 difference in the acceleration and the emission structures   
will appear as a difference in the predicted pulse profiles, 
the phase-resolved spectra and the 
polarization characteristics, which can be examined by a 
three-dimensional model.

Within the framework of the traditional outer gap model, 
  Romani \& Yadigaroglu (1995) 
considered a three-dimensional geometry  and explained 
the general features of the observed  
pulse profile such as two-peaks in a single period. 
 Subsequently, Cheng et al. (2000, hereafter CRZ00) 
developed the three-dimensional outer gap model, in which the gap is 
 sustained self-consistently  by the pair-creation process between the primary 
photons emitted via the curvature process and the surface 
$X$-ray photons. CRZ00 calculated the phase-resolved 
spectrum in $\gamma$-ray region for the Crab pulsar with 
the synchrotron radiation and 
the inverse Compton scattering  of the electron and positron pairs produced 
outside the gap.  Zhang \& Cheng (2002) 
reconsidered the CRZ00 model to examine the phase-resolved spectra in 
$X$-ray region. However, it has been difficult to 
explain the detail features of the 
observed  pulse profiles and phase-resolved spectra
 with the traditional model. 
Furthermore, the previous studies 
have not discussed  the complex features of the observed pulse profiles from 
optical to $\gamma$-ray bands. 
Recently, Takata et al. (2007) explained  the observed 
polarization characteristics in the optical wavelengths (Kanbach et al 2005)
 with the new outer gap geometry. Jia et al. (2007) examined 
the phase-resolved spectra by taking account of the emissions below 
null charge surface. However, these studies also did
 not consider the pulse profile, the phase-resolved spectra and
 the polarization characteristics in optical to $\gamma$-ray 
bands, simultaneously.

In this paper, 
we  study the emission process of the Crab-like pulsars with the outer
 gap accelerator model from both   
a two-dimensional electrodynamical model and a three-dimensional 
emission model point of views. In first part (section~\ref{dynamics}) 
of this paper, 
we will summarize the results 
of the two-dimensional electrodynamical study, in which the outer gap
structure for the Crab pulsar is 
solved with the Poisson equation, the particle motion, the radiation process 
and the pair-creation process in meridional plane, following Takata
et al (2004, 2006) and Hirotani (2006). We will show a 
result which has a consistent GeV spectrum  with the observed
phase-averaged spectrum  of the
Crab pulsar. 
In the second part (sections~\ref{model} and \ref{results}), we will 
 conduct a three-dimensional outer gap 
model based on the results of the two-dimensional electrodynamical study. 
In the three-dimensional study, the main purpose  is 
to discuss the emission process of 
 optical to $\gamma$-ray photons  by  examining 
 the morphology change of the pulse profile 
as  a function of the photon energy and the phase-resolved spectra 
for the Crab pulsar with the outer gap accelerator model. 
We will predict the polarization characteristics through  optical 
 to $\gamma$-ray bands. We also apply the model to a Crab-like pulsar,
PSR B0540-69. The Crab pulsar and PSR B0540-69 are sometimes called 
twin pulsars, because their pulsar parameters are very similar to each other. 
However, the observed shapes of pulse profiles are very different to 
each other. This pair will give an unique opportunity to examine 
the model capability. 

Important differences between present  
and  previous three-dimensional studies are as follows. 
First, we take into account 
the emissions both below and beyond 
the null surface as the electrodynamical study has predicted,
 while only the emissions beyond 
the null surface were taken into account in CRZ00.
 Secondary we  discuss the morphology change 
of the pules profile by calculating 
 local  emissivity as a function of the photon energy, while the 
previous studies did not discuss the morphology change because 
they assumed  a constant emissivity when  the pulse profiles were calculated. 
We deal the gyration motion of the pairs because 
the gyration motion causes the depolarization for the synchrotron radiation. 
We adopt the rotating dipole field in the observer frame, while the previous 
studies adopted it in the co-rotating frame. 
 Though these effects were considered in 
Takata et al. (2007),  they calculated only 
the synchrotron emission process and presented   
 the phase-averaged spectrum below MeV energy.  
In this paper, we extend the model spectrum up to  
$\gamma$-ray bands by computing also  the 
inverse Compton scattering. 
Finally, we calculate the collision angle of the inverse Compton
scattering between the pairs and the background synchrotron photons by tracing 
the three-dimensional trajectory of the synchrotron photons,
 while the isotropic distribution of the back ground photons was assumed 
in the previous studies (CRZ00). 
 The collision angle greatly affects to the emissivity
 of  and the polarization characteristics of the inverse Compton scattering. 
By including all these effects, we examine the pulse profiles, the 
phase-resolved spectra and the polarization characteristics in
 optical to $\gamma$-ray bands, simultaneously.

\section{Results of Two-dimensional Electrodynamical Model}
\label{dynamics}
In this section, we summarize the results of the two-dimensional 
electrodynamical study for the Crab pulsar. 
Following Takata et al (2004, 2006) and Hirotani (2006) 
we calculate the spectrum of 
the synchrotron and curvature radiation processes of the primary particles 
with the electric structure  by solving 
 the Poisson equation [$\nabla^2\Phi=-4\pi (\rho-\rho_{GJ})$], 
the equation  of motion for the particles, 
 the pair-creation process and the radiation process.  
As discussed in Takata el al. (2004, 2006), the electric structure depends 
on the current and the gap size, which are model parameters in their studies. 
In this section,  we show a result, which produces a consistent
 GeV spectrum with the observations. We ignore the effect of the gravity 
which is not important for the dynamics of the outer gap accelerator. 
We  adopt static dipole field, while in the later section of 
the three-dimensional study, we apply the rotating dipole field. 
The obtained electric structure with the static and the rotating dipole field
  did not change  very much, because the radial distances to the null 
charge points, that is, to the gap position  are similar to each other. 
For the pair-creation process in the
 gap, we consider the thermal soft-photons coming from the stellar surface.
 We adopt $kT=170$~eV for the Crab pulsar (Yakovlev \& Pethick 2004). 
The inclination angle is assumed as 
$\alpha=50^{\circ}$.

\begin{figure}
\epsscale{.80}
\plotone{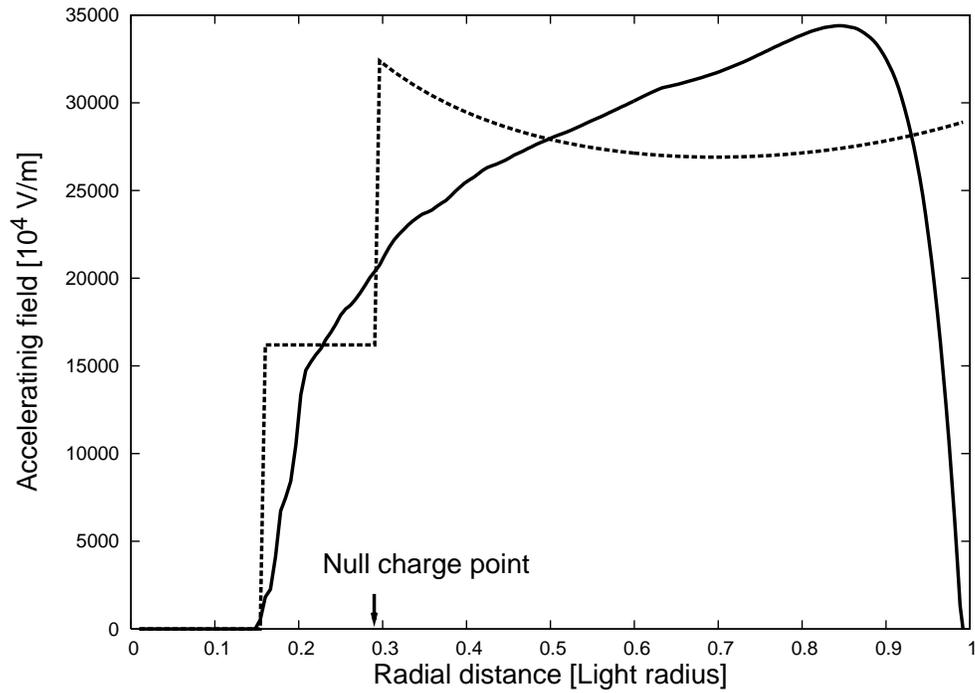}
\caption{Electric structure in the gap. The solid line shows 
the solved accelerating field along the field line, which locate at 50~\% of 
the thickness from the last open field line. This results are for 
The inclination angle is $50^{\circ}$ and for about 22~\% of the 
Goldreich-Julian current in the gap. 
Beyond the null charge point, the dashed line is vacuum electric field 
obtained by Cheng et al (1986a).   
\label{Elect}}
\end{figure}

Thick solid line in Figure~\ref{Elect}  shows the solved 
accelerating electric field along the field line locating at 50~\% 
of the trans-field thickness from the lower boundary (last-open field line). 
Here, we assume $0.1R_{lc}$ of the gap thickness at the light cylinder and the 
outer boundary is putted at near the light cylinder. We also assume that $5\%$
 of the Goldreich-Julian current, $0.05\Omega B/2\pi$, is injected 
at the outer boundary. 

The position of the inner boundary is solved with the current, 
for which about 22~\% 
of the Goldreich-Julian current runs through the outer gap
 in the present case. 
From Figure~\ref{Elect}, we can see that the inner boundary 
($r\sim 0.18R_{lc}$) is inside of the null charge point ($r\sim 0.29R_{lc}$). 
As suggested by Takata et al. (2004), the inner boundary is located 
at the position, on which $j_g+j_2-j_1\sim B_z/B$ is satisfies, where $j_g$ 
is non-dimensional current created in the gap, $j_1$ and $j_2$ 
are non-dimensional current injected at the inner and 
the outer boundaries, respectively. For example, 
for  no injection currents, $j_1=j_2=0$, the inner boundary is located at 
the null charge surface, where $B_z=0$, 
 if no current is created inside of the gap ($j_g=0$) as the vacuum case.
On the other hands, if $j_g\sim\cos\alpha$, where $\alpha$
 is the inclination angle, is created, the inner boundary 
is located at the stellar surface on which $B_z/B\sim\cos\alpha$ 
is satisfied around the magnetic pole. In the present case, 
the inner boundary is located at the position of 
about 65~\% of the radial distance to the null point 
with the current components $(j_g,~j_1,~j_2)=(0.17,~0,~0.05)$. 

\begin{figure}
\epsscale{.80}
\plotone{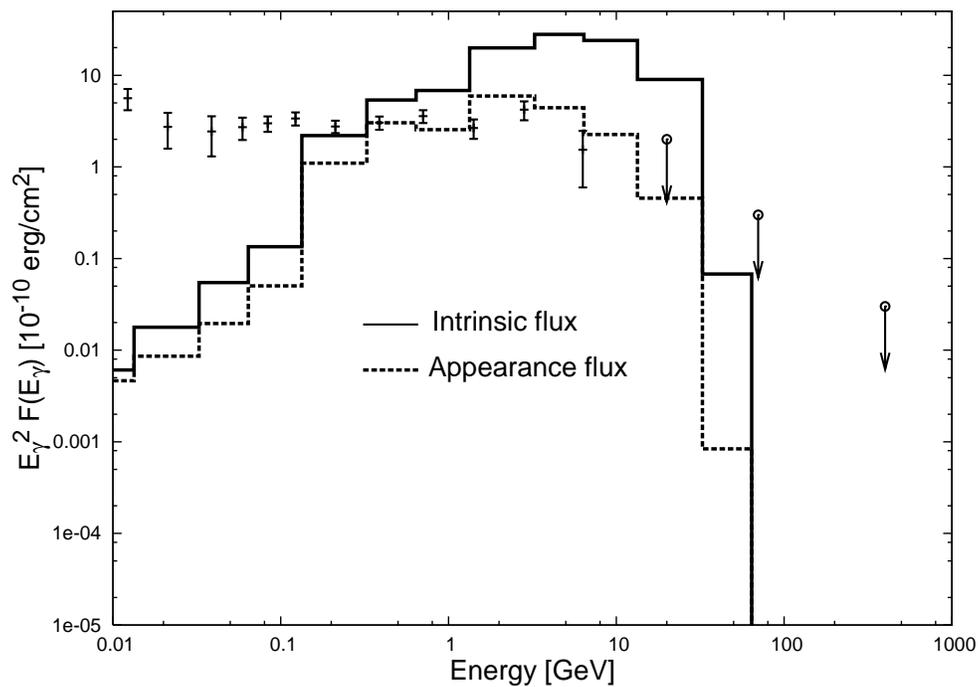}
\caption{The spectrum of the synchrotron and the curvature processes 
of the primary particles in the outer gap. The solid line represent 
the intrinsic flux from the outer gap. The dashed line show the appearance 
flux after pair-creation process outside of the outer gap with 
the soft photons emitted by the synchrotron radiation of the secondary pairs. 
The observation data are taken from Kuiper et al. (2002) and reference 
therein.  \label{spectrum}}
\end{figure}

Figure~\ref{spectrum}  shows the calculated synchrotron-curvature
spectrum and compares with the observed phase-averaged spectrum.  
Sold-line shows 
spectrum of the intrinsic radiation from the outer gap, while the dashed-line 
represents the appearance spectrum after attenuation  
of the photons via pair-creation 
process outside of the gap with the soft photon-field emitted 
by the synchrotron process of the secondary pairs.  We calculate 
the initial pitch angle of the pairs from  the propagating direction of 
the curvature photons  and the magnetic field direction at the pair-creation 
position.  For obtaining
 the luminosity, we assume the gap opening angle $\sim 250$~degree in 
the azimuthal direction (see section~\ref{efield}). 

As dashed-line shows the large amount of the curvature photons above 500~MeV 
are converted into the secondary pairs outside of the gap
 via the pair creation process with the X-ray photons from the secondary pairs.
  We find from Figure~\ref{spectrum} 
 that the shape of the spectral energy distribution 
after absorption becomes relatively flat and  explains the observation 
above  $100$~MeV. 

The inverse-Compton process of the primary particles 
in the gap is a  possible mechanism for TeV emissions. However, 
the present model predicts the TeV flux for the Crab pulsar 
is too low to detect the present Cherenkov telescopes. The soft-photons
 emitted by secondary pairs above the gap may not be able to illuminate  
the gap due to the curvature of the field lines, and only thermal photons from 
the stellar surface may be scattered by the primary particles. In such a case, 
we found that the intrinsic flux on the Earth becomes 
$\sim 10^{-15}~\mathrm{erg/cm^2 s}$~, which is much small
 compared with the sensitivity of the present Cherenkov telescope. 
 On the other hand,  some soft-photons emitted by secondary pairs
 may illuminate the outer gap because of the effects of the pith angle. 
In such a case, we obtained  intrinsic TeV flux
 which is easily detected by present instruments. 
However, 
the optical depth of the pair-creation for TeV photons is much larger than 
unity in the magnetosphere, and the residual TeV photons 
 are a very few, which is difficult to detect with 
the present instruments.  

From  Figure~\ref{spectrum}, we can see that the synchrotron and 
curvature radiations of 
the primary particles in the outer-gap does not explain the observed flux 
below 100~MeV. We consider that the secondary pairs created outside gap produce
 below 100~MeV photons via the synchrotron and the inverse-Compton process. 
Furthermore, the present two-dimensional model can compare with only the 
phase-averaged spectrum. More detailed observation such like
 the pulse profile, the phase-resolved spectra and the polarization  require a
 three-dimensional model.  
Following sections, therefore, we calculate the emission process 
of the secondary pairs and conduct  a three-dimensional model. 

\section{A Three-Dimensional Emission  Model}
\label{model}
In the following, we conduct a three-dimensional emission model. 
We anticipate that the emission direction is coincide with the particle
 motion in the observer frame. 
In the present paper, we adopt 
the rotating dipole field in the observer frame while it was assumed 
in the co-rotating frame in the previous studies (Romani \& Yadigaroglu 1995; 
Cheng et al 2000; Dyks et al 2004). 
As a results, the magnetic 
field configurations and resultant the morphology of emission pattern 
 in the observer frame 
are different between the present and the previous studies, but 
 the difference becomes  to be important only near the light cylinder. 
 In the present study, furthermore, 
we discuss the model in the observer frame only, and 
we do not introduce the co-rotating frame.

\subsection{Electric field}
\label{efield}
We have to describe the accelerating electric field into the
three-dimensional from. Based on the result (Figure~\ref{Elect}) of
the two-dimensional electrodynamical study,  we adopt the following
 three-dimensional form. 
First,  we use the vacuum solution obtained 
by Cheng et al. (1986a),   
\begin{equation}
E_{||}(r)=\frac{\Omega B(r)f^2(r)R_{lc}^2}{c s(r)}, 
\label{electric}
\end{equation} 
beyond the null charge surface, 
where $s(r)$ is the curvature radius of the magnetic field line and $f(r)$
 is the fractional gap thickness.  We can calculate the electric field 
at each point having a three-dimensional radial distance, $r$.
  Below null surface,
  we  assume 50~\% of the strength of the electric 
field at the null point given by equation~(\ref{electric}). 
The dashed-line 
in Figure~\ref{Elect} shows the electric field strength of the approximation 
form  in the meridional plane. 
We can see that  the 
typical strength of the  electric field by the
two-dimensional electrodynamical study (solid-line) in the meridional
plane  is in general described  by the present simple form (dashed-line). 
In fact,  as long as the 
gap is geometrically
thin in the trans-field direction and the magnitude of the current is smaller 
than the Goldreich-Julian value, the vacuum solution beyond the null
charge surface  approximately describes the typical strength of the
accelerating field in the meridional plane.  Now, we assume that this
simple form can  describe also 
 the typical strength of the three-dimensional distribution of the
accelerating  electric field.

At each point, the maximum Lorentz factor of the particles are determined 
by the force balance between the acceleration  by the electric 
 and  curvature radiation back reaction,  
$\Gamma_p(r)=[3s^2(r)E_{||}/2e]^{1/4}$, where 
$\Omega_2=\Omega/\mathrm{100s^{-1}}$ . 
The primary particle emits high-energy photons as the curvature 
radiation process, whose  typical energy  is  
$E_{curv}(r)=3h\Gamma_{p}^3(r)c/4\pi s(r)$,
 and the local power of the curvature
radiation is given by $l_{curv}=eE_{||}c$. 

It is important to estimate the polar cap opening angle of the active 
region of the outer gap accelerator. When we consider an open-field
 line through the outer gap, 
 the pair-creation process between the primary curvature photons and 
the surface $X$-rays mainly occurs near and below  
the null charge surface. Therefore, we may be able to
 relate the opening angle with the pair-creation mean free path,
 which is estimated as $l(r)\sim[2s(r)f(r)R_{lc}]\sim 2 f^{1/2}(R_{lc}/2)r$,  
at the null surface. 
In the present paper, we constrain the width of the polar cap
 angle of the active gap by the condition that  
the mean-free path at the null charge point on the magnetic field line 
 becomes shorter than 
the light radius. This condition produces 
the width of the polar cap  angle of $\sim 250^{\circ}$.

\subsection{Distribution and motion of the secondary pairs}
\label{secondary}
As we demonstrated in section~\ref{dynamics},  a significant amount of
 the curvature photons 
above $\sim$500~MeV convert into secondary pairs outside  the gap via
 the photon-photon pair-creation process with the soft-photons emitted 
by the synchrotron radiation of the secondary pairs. 
 From Figure~\ref{spectrum}, furthermore, 
 we can read  that the spectrum of the intrinsic emissions has 
 the photon index of about $-1$. 
Therefore, we may approximately describe the local curvature spectrum with   
 $F_{curv}\sim l_{curv}j n_{GJ}/E_{curv}E_{\gamma}$, where $j$ 
represent the current in units of the Goldreich-Julian value. 
Using the steady loss equation, 
$d[\dot{E}_edn/dE_e]/dE_e=Q(E_e)$,  we obtain  
the distribution of the secondary pairs as  
\begin{equation}
\frac{dn_e}{dE_{e}}\sim\left\{
\begin{array}{@{\,}ll}
l_{curv}j n_{GJ}\mathrm{ln}(E_{curv}/2E_e)/\dot{E}_eE_{curv}
& \mathrm{for}~~500~\mathrm{MeV}<2E_e<E_{curv} \\
l_{curv}j n_{GJ}\mathrm{ln}(E_{curv}/500~\mathrm{MeV})/\dot{E}_eE_{curv}
& \mathrm{for}~~2E_e<500~\mathrm{MeV} 
\end{array}
\right.
\label{dist}
\end{equation}
where  $\dot{E}_{e}= 2e^4B^2(r)\sin^2\theta_p(r)
\Gamma_e^2/3m_e^2c^3$ is the energy loss rate of the synchrotron
radiation of the secondary pairs, $\theta_p$ is the pitch angle, 
and $\Gamma_e$ is the Lorentz 
factor of the pairs. The local pitch angle will be expressed as  
$\sin\theta_{p}(r)\propto\sqrt{2f(r)R_{L}/s(r)}$. Because 
$f(r)=f(R_{lc})(r/R_{lc})^{3/2}$ and $s(r)\sim\sqrt{rR_{lc}}$ are 
 satisfied for the dipole field, we may relate with   
$\sin\theta_{p}(r)=(r/R_{lc})^{1/2}\sin\theta_{p}(R_{lc})$ between the
 pitch angle of the local point and the light cylinder along the field lines. 
 Outside the gap, the pairs loose  most of their energy via 
the synchrotron process. Because the synchrotron loss rate $\dot{E}_e$ is 
proportional to square of the particle energy, the power law index of
the distribution given by equation (\ref{dist}) becomes $p\sim2$, 
which produces a synchrotron spectrum with the spectral index of 
$\alpha_{\nu}\sim0.5$.

In the observer frame, we may describe the particle motion outside the gap 
 with 
\begin{equation}
\mbox{\boldmath$\beta$}=\beta_0\cos\theta_p\mbox{\boldmath$b$}
+\beta_0\sin\theta_p\mbox{\boldmath$b$}_
{\perp}+\beta_{co}\mbox{\boldmath$e$}_{\phi}, 
\label{pmotion}
\end{equation}
where the first term and the second term 
in the right hand side represent,
 respectively, the particle motion parallel to the magnetic field 
and gyration motion, and the third term represents 
the co-rotational motion with the non-dimensional 
velocity $\beta_{co}=\varpi/R_{lc}$, where $\varpi$ is 
the axial distance.   The  vector $\mbox{\boldmath$b$}$
 is the  unit vector along the field line and $\mbox{\boldmath$b$}_{\perp}$ 
represents the  unit vector perpendicular to the magnetic field line,
$\mbox{\boldmath$b$}_{\perp}\equiv\pm(\cos\delta\phi\mbox{\boldmath$K$}
+\sin\delta\phi\mbox{\boldmath$K$}
\times\mbox{\boldmath$b$})$, where the sign $+$ (or $-$) corresponds
to gyration of  the positrons (or electrons), 
$\mbox{\boldmath$K$}=(\mbox{\boldmath$b$}\cdot\nabla)\mbox{\boldmath$b$}/
|(\mbox{\boldmath$b$}\cdot\nabla)\mbox{\boldmath$b$}|$ is 
the unit vector of the curvature of the magnetic field line, and 
$\delta\phi$ represents the phase of gyration around the magnetic field.
Because  the pairs have an ultra-relativistic speed, we determine
the value of the coefficient $\beta_0$ from the condition 
that $|\mbox{\boldmath$\beta$}|=1$. We anticipate that the photons are
 emitted in the direction of the particle motion of equation~(\ref{pmotion}).

We note that the synchrotron radiation after collecting of the photons 
 is greatly depolarized due to the gyration motion of the pairs, 
 although the intrinsic radiation is highly polarized. 
Therefore, the observed small 
polarization degree $\sim10~\%$  at the optical bands 
for not only the Crab pulsar, but also for the Vela pulsar 
(Mignami et al. 2007) are easily reproduced by the synchrotron 
emission model (Takata et al. 2007).

We assume that the emission region of the secondary pairs extends 
just above the outer gap with thickness of the mean free path of the 
pair-creation  $\lambda\sim 10^{7}$~cm$\sim 0.1R_{lc}$ for the Crab pulsar. 
Some secondary high-energy 
photons via the   inverse Compton scattering  
may convert into the tertiary pairs. The 
tertiary pairs  will 
be produced above the emission region of the secondary pairs, 
and its initial Lorentz factor will be smaller than that of 
the secondary pairs.  We also 
take into account the effects of the emissions from the tertiary pairs. 
 
\subsection{Emission process of  the secondary pairs and polarization}
We consider that the synchrotron radiation and 
the inverse Compton scattering of the secondary pairs 
are major emission mechanisms
 for the observed non-thermal radiation through optical to $\gamma$-ray
bands for the Crab-like pulsars. If we estimate  the radiation powers 
of the synchrotron radiation and the inverse Compton scattering, we obtain
\begin{equation}
\frac{P_{syn}}{P_{IC}}\sim 10\left(\frac{U_{ph}}{5\cdot10^{7}\mathrm{erg/cm^3}}
\right)^{-1}\left(\frac{B}{10^{6}\mathrm{Gauss}}\right)^2
\left(\frac{\sin\theta_p}{0.1}\right)^2
\label{fratio}
\end{equation}
with $U_{ph}$  being the energy density of the synchrotron photons.
 The estimated value will explain the observed flux ratio
 of 1~MeV and 100~MeV emissions of the Crab pulsar. 
 We calculate  only the outward emissions,
 because the inward emissions 
are expected to be much fainter than the outward emissions.

In the calculation,  
we firstly compute the volume emissivity of the 
synchrotron radiation and its emitting  direction for each radiating point 
(section~\ref{synch}). 
Then, we trace the propagation of the synchrotron 
beam to simulate the  scattering process by  the pairs (section~\ref{inv}). 
On each scattering point, 
we calculate the volume emissivity and the polarization of 
the inverse Compton scattering for 
 a specific viewing angle of the observer. We perform this procedure 
for all calculation points to obtain the  total
radiation for the specific observer.  
This procedure is equivalent with computing  the radiation transfer,  
\begin{equation}
\frac{dI(\mbox{\boldmath$k$}_1,\epsilon_1)}{ds}
=j_{s}(\mbox{\boldmath$k$}_1,\epsilon_1)
+j_i(\mbox{\boldmath$k$}_1,\epsilon_1),
\end{equation}
where $I(\mbox{\boldmath$k$}_1, \epsilon_1)$ is the total intensity 
of the beam 
 propagating in the direction of $\mbox{\boldmath$k$}_1$, $\epsilon_1$ is 
 the energy of photons in units of the electron rest mass energy, 
$j_s(\mbox{\boldmath$k$}_1, \epsilon_1)$ 
is the volume emissivity of the synchrotron radiation, and 
$j_i(\mbox{\boldmath$k$}_1,\epsilon_1)$ 
represents the amount of the scattered photons into the direction 
$\mbox{\boldmath$k$}_1$  and to the energy $\epsilon_1$.  
We neglect the effects of the absorption, 
because the synchrotron self-absorption is not important above optical 
photon energy, where we are now interested in.  
Also,  we ignore the effects of the scattering off from the direction 
$\mbox{\boldmath$k$}_1$ of the synchrotron photons,
 because the scattered photons are  
tiny amounts of total number of the synchrotron photons.

\subsubsection{Synchrotron radiation}
\label{synch}
Assuming that all the synchrotron photons are radiated toward 
 the particle motion direction, $\mbox{\boldmath$k$}_1=
\mbox{\boldmath$\beta$}$, 
the volume emissivity of the synchrotron radiation is 
calculated from 
\begin{equation}
j_s(\epsilon_1)\equiv\frac{dI_s}{ds}=\frac{\epsilon_1
F_{syn}(\epsilon_1)}
{\delta\Omega},
\label{volu}
\end{equation}
 where $\delta\Omega$ is the solid angle of the radiation,  and $F_{syn}$ is
 the photon spectrum described by 
 \begin{equation}
F_{syn}(\epsilon_1)
=\frac{3^{1/2}e^3B(r)\sin\theta_{p}(r)}{mc^2h
\epsilon_1}\int\left[\frac{dn_e(r)}{dE_e}\right]
F(\epsilon_1/\epsilon_{syn})dE_e,
\label{emis}
 \end{equation}
where $\epsilon_{syn}(r)=3he\Gamma_e^2(r)B(r)\sin\theta_p(r)/4\pi m^2_ec^3$ 
is the typical photon energy of the
 pairs in units of the electron rest mass energy, 
$\Gamma_e$ represents the Lorentz factor of the secondary 
pairs and 
 $F(x)=x\int_x^{\infty} K_{5/3}(y)dy$ with $K_{5/3}$ being the modified
Bessel function of order 5/3. 
 
When we calculate the polarization of the synchrotron radiation, 
we anticipate that  direction of the 
electric vector of the electro-magnetic wave propagating toward the observer 
is parallel to the projected direction of the acceleration 
of the particle on the sky, $\mbox{\boldmath$E$}_{em}
\propto \mbox{\boldmath$a$}
-(\mbox{\boldmath$k$}_1\cdot\mbox{\boldmath$a$})\mbox{\boldmath$k$}_1$,
(Blaskiewicz et al. 1991), where the acceleration vector $\mbox{\boldmath$a$}$ 
derived from equation (\ref{pmotion}) 
is approximately written by $\mbox{\boldmath$a$}
\sim \pm\beta_{0}\omega_B\sin\theta_p(-\sin\delta\phi\mbox{\boldmath$K$}
+\cos\delta\phi\mbox{\boldmath$K$}\times\mbox{\boldmath$b$})$, where $\omega_B$
 is the gyration frequency. 
We assume that the radiation at each point 
is linearly polarized with degree of $\Pi_{syn}=(p+1)/(p+7/2)$, where $p$ is 
the power law index of the particle distribution. Because 
the observed radiation is consist of the radiations from the different 
particles with the different pitch angle, we assume that 
the circular polarization will cancel out and become zero 
in the observed radiation. 
The Stokes parameters 
$Q_{syn}$ and $U_{syn}$ are, 
respectively, calculated from  
$dQ_{syn}(\mbox{\boldmath$k$}_1,\epsilon_1)/ds
=j_s(\mbox{\boldmath$k$}_1,\epsilon_1)\cos2\eta_s(r)$ and
 $dU_{syn}(\mbox{\boldmath$k$}_1,\epsilon_1)/ds
=j_{s}(\mbox{\boldmath$k$}_1,\epsilon_1)\sin2\eta_s(r)$, 
where $\eta_s(r)$ is 
the position angle defined by the angle between 
the electric vector of the wave and the projected direction of the rotation 
axis on the sky, 
 $\mbox{\boldmath$\Omega$}_p=\mbox{\boldmath$\Omega$}-
(\mbox{\boldmath$k$}_1\cdot\mbox{\boldmath$\Omega$})\mbox{\boldmath$k$}_1$.

\subsubsection{Inverse Compton scattering}
\label{inv}
 To simulate the scattering process, 
we trace the three-dimensional trajectory 
of the synchrotron photons. 
When we trace the trajectory of the synchrotron photons, 
we define the  Cartesian coordinate such that $z$-axis is along the rotation 
axis and the $x$-axis is in  the meridional plane. 
 By  ignoring  bending of the trajectory 
due to the gravity, the position of the photons after traveling distance 
$\delta s$ is  $x(\delta s)=(x_0+k_{x_0}\delta s)\cos(\delta s/R_{lc})
+(y_0+k_{y_0}\delta s)\sin(\delta s/R_{lc})$, 
$y(\delta s)=-(x_0+k_{x_0}\delta s)\sin(\delta s/R_{lc})+
(y_0+k_{y_0}\delta s)\cos(\delta s/R_{lc})$ 
and $z(\delta s)=z_0+k_{z_0}\delta s$, where the coordinates 
$(x_0,~y_0,~z_0)$ are the radiating point of the synchrotron photon, and 
$(k_{x_0},~k_{y_0},~k_{z_0})$ represents the emission direction 
at $(x_0,~y_0,~z_0)$. The emission direction of the background synchrotron 
radiation is calculated from equation~\ref{pmotion}, which takes account 
the aberration due to the corotating motion. 

Because the mean free path of a synchrotron photon of the scattering
 is much longer than the light radius, one can 
consider that the scattering rate 
is constant along the path of the synchrotron photons  
 in the magnetosphere as the first order approximation.
 We determine 
the scattering points at regular interval, which is 
much shorter than the gap size, along the path of 
 the synchrotron photons. 
In the calculation, we first compute the Stokes parameter 
of the Compton process in the electron rest frame, 
and then we transform it  to the observer frame.
 In the following, 
the prime and 'non'-prime quantities represent the quantities in the electron 
rest frame and the observer frame, respectively. A detail derivation 
of equations of the 
Stokes parameters (\ref{unpst}) and (\ref{pst})  are seen 
 in Appendix~\ref{appen}.

We denote the specific intensity of the synchrotron 
radiation propagating to the direction $\mbox{\boldmath$k$}_0$ in the 
observer frame with 
$I_0(\mbox{\boldmath$k$}_0,\epsilon_0)$, where $\epsilon_0$ 
represents the energy of the background photons in units of the electron 
rest mass energy. In the electron rest frame, the background radiation becomes 
$I'_0(\mbox{\boldmath$k$}'_0,\epsilon'_0)=D_1^3
I_0(\mbox{\boldmath$k$}_0,\epsilon_0)$, where $D_1=\epsilon_0'/\epsilon_0=
\Gamma_e^{-1}(1+\beta\cos\theta'_0) $ is a Doppler factor, $\beta$ 
is the velocity of the scattering particles in units of the speed of light, 
 and $\Gamma_e=1/\sqrt{1-\beta^2}$.
 The polar angle $\theta_0$ is defined by 
the angle between the directions of the particle motion and  
of the propagation of 
the background radiation, which becomes $\theta_0\sim 0.1-0.3~$radian 
in numerically.  For the particles with the Lorentz 
factor $10^3\sim 10^4$, 
optical to $X$-ray photons  are mainly scattered. 
In this photon energy bands, 
the  synchrotron photons are distributed 
with a spectral index of $\alpha_{\nu}\sim0.5$ because the cut-off energy 
 of the synchrotron spectrum is $\sim 1$~MeV and because 
 the particles are  distributed with the index  
$p\sim2$ (section~\ref{secondary}).
Because the synchrotron beam from each position is strongly collimated, 
we approximately describe 
the background beam,  in which  the center of the beam 
is direct to  the polar angle 
$\theta_0$ measured from the electron motion direction and the azimuthal 
direction $\phi_0$, as  $I_0(\mbox{\boldmath$k$}_0,\epsilon_)
=C_0\epsilon^{-0.5}\delta(\theta-\theta_0)\delta(\phi-\phi_0)$, where $C_0$ 
is evaluated from equation (\ref{volu}).

We are interested in the inverse Compton scattering 
with the background  synchrotron 
radiation,  which is partially polarized
 with $\Pi_{syn}\sim 70\%$.  
With unpolarized components of the background radiation propagating  to  
 the direction $\mbox{\boldmath$k$}_0$, 
the volume emissivity $j_{u}$ and the Stoke parameter $Q_{u}$ and $U_{u}$ 
of the scattered radiation propagating  to the direction 
 $\mbox{\boldmath$k$}_1$ are calculated from (see appendix~\ref{appen})
\begin{eqnarray}
&&\left.
\begin{array}{@{\,}ll}
&j_{u}(\mbox{\boldmath$k$}_1,\epsilon_1)\equiv dI_{u}
(\mbox{\boldmath$k$}_1,\epsilon_1)/ds \\
&dQ_{u}(\mbox{\boldmath$k$}_1,\epsilon_1)/ds \\
&dU_{u}(\mbox{\boldmath$k$}_1,\epsilon_1)/ds
\end{array}
\right\}=(1-\Pi_{syn})
\frac{3\sigma_T}{16\pi}C_0\int d\Gamma_e 
\left[\frac{dn_e}{d\Gamma_e}\right] \nonumber \\
&\times&
\frac{\epsilon^{'-a}_0}{\Gamma_{e}^{4+a}(1-\beta\cos\theta_1)^2
(1+\beta\cos\theta'_0)^{a+1}}\left(\frac{\epsilon'_1}{\epsilon'_0}\right)^2 
\left\{
\begin{array}{@{\,}ll}
\left[\frac{\epsilon'_0}{\epsilon'_1}+\frac{\epsilon'_1}{\epsilon'_0}
-\sin^2w'_s\right],& \\
q'_u\cos2\zeta-u'_u\sin2\zeta,& \\
q'_u\sin2\zeta+u'_u\cos2\zeta,&
\end{array}
\right.
\label{unpst}
\end{eqnarray}
and 
\begin{eqnarray}
q'_u&=&\sin^2w'_s\cos2\eta', \nonumber \\
u'_u&=&\sin^2w'_s\sin2\eta',
\end{eqnarray}
where the Stokes parameters are measured from the rotation axis of the 
pulsar projected on the sky, and $\zeta$ is defined by the angle between 
the directions of the  rotation axis and the particle motion projected  
on the sky (Figures~\ref{coord}). The polar angle 
$\theta_1$ represents the propagating direction of the scattered photons 
 measured  from the particle motion direction, $w_s$ is the scattering 
angle defined by 
$\cos w_s=\mbox{\boldmath$k$}_0\cdot\mbox{\boldmath$k$}_1$ and
 the azimuthal angle $\eta$ is the angle between the orthogonal direction to 
the scattering plane and the direction of the particle motion projected 
on the sky. 

For the polarized component of the background radiation, 
the volume emissivity and the 
Stokes parameters are calculated from   
\begin{eqnarray}
&&
\left.
\begin{array}{@{\,}ll}
&j_{p}(\mbox{\boldmath$k$}_1,\epsilon_1)\equiv
dI_{p}(\mbox{\boldmath$k$}_1,\epsilon_1)/ds \\
&dQ_{p}(\mbox{\boldmath$k$}_1,\epsilon_1)/ds \\
&dU_{p}(\mbox{\boldmath$k$}_1,\epsilon_1)/ds
\end{array}
\right\}=\Pi_{syn}
\frac{3\sigma_T}{16\pi}C_0\int d\Gamma_e 
\left[\frac{dn_e}{d\Gamma_e}\right]
\nonumber \\
&\times&
\frac{\epsilon^{'-a}_0}{\Gamma_{e}^{4+a}(1-\beta\cos\theta_1)^2
(1+\beta\cos\theta'_0)^{a+1}}\left(\frac{\epsilon'_1}{\epsilon'_0}\right)^2 
\left\{
\begin{array}{@{\,}ll}
\left[\frac{\epsilon'_0}{\epsilon'_1}+\frac{\epsilon'_1}{\epsilon'_0}
-\sin^2w'_s\cos^2\lambda'_p\right],& \\
q'_p\cos2\zeta-u,_p\sin2\zeta& \\
q'_p\sin2\zeta+u'_p\cos2\zeta,&
\end{array}
\right.
\label{pst}
\end{eqnarray}
and 
\begin{eqnarray}
q'_p&=&\left[\sin^2w'_s-(1+\cos^2w'_s)\cos2\lambda'_p\right]
\cos2\eta'-2\cos w'_s\sin2\lambda_p'\sin2\eta' , \nonumber \\
u'_p&=&\left[\sin^2w'_s-(1+\cos^2w'_s)\cos2\lambda'_p\right]
\sin2\eta'-2\cos w'_s\sin2\lambda_p'\cos2e\eta',
\end{eqnarray}
where $\lambda_p$ is  the angle between the polarization 
plane of the background radiation and the plane of the scattering.
 By exploring the additive property of the Stokes parameters, 
the total volume emissivity and Stokes parameters are,  respectively, 
 given by $j_{i}=j_{u}+j_{p}$,  $dQ_{i}/ds=dQ_{u}/ds+dQ_{p}/ds$ and 
$dU_{i}/ds=dU_{u}/ds+dU_{p}/ds$.

After collecting all photons from the possible points for each rotation phase
 $\Phi$ and a viewing angle $\xi$, the degree of the radiation 
and the position angle of the electric vector of the radiation 
 are, respectively,  calculated from 
$P(\xi,\Phi,\epsilon_1)=\sqrt{Q^2(\xi,\Phi,\epsilon_1)
+U^2(\xi,\Phi,\epsilon_1)}/I(\xi,\Phi,\epsilon_1)$ and 
$\chi(\xi,\Phi, \epsilon_1)=0.5\mathrm{atan}[U(\xi,\Phi,\epsilon_1)/Q(\xi,\Phi
,\epsilon_1)]$,
 where $I(\xi,\Phi,\epsilon_1)$, $Q(\xi,\Phi,\epsilon_1)$ and 
$U(\xi,\Phi,\epsilon_1)$ 
are the Stokes parameters after collecting photons emitted via  
both synchrotron radiation and the inverse Compton scattering. 
The position angle 
$\chi(\xi, \Phi,\epsilon_1)$ is measured anticlockwise from the axis of the 
rotation projected on the sky (Figure~\ref{coord}). 

\subsection{Model parameter}
The inclination angle of the pulsars has been constrained by 
the polarization measurements of the radio pulsed emissions. 
However,  it has not been strongly  constrained  
 the inclination angle for the Crab pulsar and PSR B0540-69. Therefore, 
we treat the inclination angle as a model parameter.
 The viewing angles $\xi$ of the observer 
 measured from the rotational axis is also a model parameter.
For this local model in the magnetosphere, the current should be dealt as 
a model parameter and the position of the inner boundary 
depends on the assumed current (Hirotani et al. 2003;
 Takata et al. 2004, 2006; Hirotani 2006).     
Instead of the current, however, 
 the ratio of the  radial distance to the inner boundary 
and distance to the the null surface $r_n$ is parameterized and 
is assumed to be constant for each 
field line, that is, $r_{in}(\phi)/r_{n}(\phi)$=constant, such  that 
the inner boundary locates far away from the 
stellar surface if the null surface locates far away. 
For example, from Figure~\ref{Elect}, the outer gap accelerator with 
the non-dimensional current  $j\sim 0.22$ has the inner boundary at 
 $r_{in}(\phi)/r_{\phi}(\phi)\sim0.65$ in the meridional plane. 
 The altitude of the emission region of the secondary
 pairs is also model parameter, because  the magnetic field
 will be modified by rotational and plasma effects near 
the light cylinder and because  the last-open field lines  may be
 different from the traditional magnetic file lines that
 are tangent to the light cylinder in the vacuum case. To specify 
the upper surface of the outer gap, 
it is convenient to refer to the footpoint of the
 magnetic surface on  the star and to  parameterize
 the fractional polar angle  $a_f=\theta_u/\theta_{lc}$,
 where $\theta_u$ and $\theta_{lc}$ 
are  the polar angle of the footpoints of the  magnetic
 surfaces for the  gap upper surface and  the last-open field line in the 
vacuum case, respectively.  We constrain the boundary of the radial distance 
to the emission region to $r=R_{lc}$. 
In this paper, we apply the model to the Crab pulsar in
section~\ref{crab} 
 and  to a Crab-like pulsar, PSR 0540-69,
 in section~\ref{like}.

\section{Model results}
\label{results}
\subsection{The Crab pulsar}
\label{crab}

For the Crab pulsar, we adopt the inclination angle of $\alpha=50^{\circ}$,  
the viewing angle of $\xi=100^{\circ}$ and the position of the inner boundary 
described by  $r_{in}(\phi)/r_n(\phi)=0.67$, 
which were chosen in Takata et al. (2007).   In this paper 
 we chose the fractional angle of $a_f=1$ to explain 
the phase-resolved spectra, 
that is, we assume the gap upper surface with the  magnetic field lines that 
are the conventional last-open field lines in vacuum. 
The opening angle of the active outer gap in 
the azimuthal direction is set at $\delta\phi=250^{\circ}$, which as assume in 
section~\ref{efield}. 
 
\begin{figure}
\epsscale{.80}
\plotone{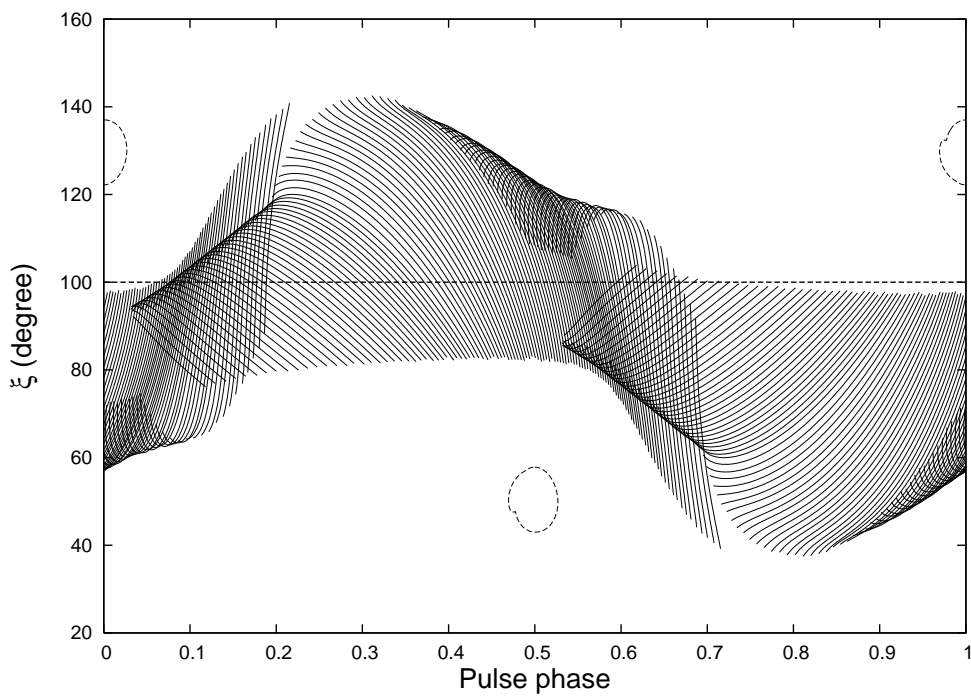}
\caption{Emission region projected onto the $(\xi,~\Phi)$-plane for
the magnetic surface $a_f=1$. Here,the emission direction
 is tangent to the direction of the local magnetic field.
 The inclination angle is
$\alpha=50^{\circ}$, the emission region extends from $r_{in}=0.67$
to $r=R_{lc}$, and the width of the polar cap angle is about $250^{\circ}$. 
The circles show the shape of the polar cap. The viewing angle 
$\xi=100^{\circ}$
(horizontal dashed-line) 
is used in the calculations for Figure~\ref{emire}$\sim$~\ref{phase50}. 
\label{map50}}
\end{figure}

Figure~\ref{map50} is the photon mapping  of the outwardly  
propagating photons, where the emission direction tangent to 
the local field lines, which is described by $a_f=1$, 
 were temporary assumed.  
Figure~\ref{emire} shows the variations of the typical 
radial distance to the emission points of the photons measured by 
the observer with the viewing angle $\xi=100^{\circ}$. 
The dashed-line in  Figure~\ref{emire}
  represents the radial distance to the emission points 
that locate beyond the null surface on the magnetic field lines  
coming from the north pole. 
And, the dotted-lines show the distance to the  points that locate
 below the null surface on the  magnetic field from the south pole. 
In the traditional study, only the emission beyond the null  surface 
(dashed-line)   have been considered. 
 We will see that 
the emission component below null surface is required 
to explain the phase-resolved spectrum of Peak~1 (Figure~\ref{phase50}). 
To calculate the phase-resolved spectra, we define the phase 
intervals of  Peak~1,  Bridge, and  Peak~2 as $0.06-0.16$, 
$0.29-0.4$, and $0.49-0.6$ (Figure~\ref{emire}).

\subsubsection{Pulse profile and Polarization}
\label{pulse}
Figures~\ref{pulsl} and~\ref{pulsh} 
show the predicted variations of the intensity (upper), 
the position angle of the electric vector of the radiation (middle)
 and the degree of the polarization (lower) 
as a function of the pulse phase from optical to $\gamma$-ray bands. 
To  compare with the observed pulse profiles in  Kuiper et al. (2001), 
 the results were calculated by integrating 
 the photons within the energy  interval $1-10$~eV, 
$0.1-2.4$~keV, $20-100$~keV, $100-315$~keV, $0.75-10$~MeV and $30-100$~MeV. 
In the figure, we define the rotation phase $\Phi=0$ in abscissa axis as lying 
the south pole, and 
the zero degree in the position angle of the bottom panel 
 is corresponding to the direction of the 
rotation axis projected on the sky. 

In Figures~\ref{pulsl} and~\ref{pulsh}, 
 we see that   the calculated 
pulse profile morphology  changes significantly as a function of
the photon energy likewise  the observational pulse profiles (see figure~5 
in Kuiper et al. 2001). One can see  in the results  that the Peak~1 emissions 
 dominate in the pulse profile 
in optical bands (left column in Figure~\ref{pulsl}).
 But,  the intensity ratio of
   Peak~1 and Peak~2 decreases with increase of the  photon energy 
in optical to  soft X-ray bands
(middle and right columns in Figure~\ref{pulsl}), and the ratio
becomes unity  around 100~keV. Eventually, the  Peak~2 emissions 
dominate the Peak~1 emissions in hard $X$-ray and soft $\gamma$-ray bands
as left and middle columns of Figure~\ref{pulsh} show.  
In the hard $\gamma$-ray bands,  furthermore, 
 Peak~1 again dominates in the pulse profile 
(right column in Figure~\ref{pulsh}). 
 This  predicted morphology change of the pulse profiles 
reproduces the observation very well.

With the present outer gap model, 
the morphology change of the pulse profiles is explained by the 
radiation properties beyond the null surface, because the  major part
of the total photons are produced there. 
The intensity of Peak~1 is much stronger  than that 
of Peak~2 in optical wavelengths,  because  pileup of the photons emitted 
from the various positions due to the special relativistic corrections 
(i.e. the aberration of the emitting direction and the flight-time) 
 occur more efficiently at the Peak~1 phase than the Peak~2 phase. 
This effect has been considered in the previous studies  
(Romani \& Yadigaroglu 1995; CRZ00) with a constant emissivity. 
 But, the present calculation produces a  Peak~2
 emission, which  becomes more and more pronounced with increasing 
energies,  and eventually dominates  the Peak~1 emission in the hard $X$-ray 
and the soft $\gamma$-ray bands. This is 
because  many  photons of  Peak~2 are 
emitted near the 
stellar surface  and on the other hand, most of
the photons for Peak~1 are emitted near the light cylinder, as
dashed-line in Figure~\ref{emire}  shows. As a result, 
 the phase-resolved spectrum of Peak~2 of the synchrotron photons 
 is harder than that of  Peak~1 so
that Peak~2 dominates in the pulse profile in hard $X$-ray and soft
$\gamma$-ray bands. And finally, the Peak~1 emissions dominate 
in the hard $\gamma$-ray region, where the inverse-Compton process is 
major emission process (Figure~\ref{phase50}).  
 For   the inverse Compton scattering  
 with the synchrotron background photons,  
the averaged strength of the magnetic 
field affects the power of radiation following equation~(\ref{fratio}). 
Because the averaged  magnetic field strength in the emission region for Peak~1
 is smaller than that for  Peak~2, 
the power of the inverse Compton scattering 
 in the emission region in  Peak~1 is larger  than that in Peak~2,
 and as a result  Peak~1 dominates in the pulse profile in 30-100~MeV bands.
  Thus, we find that  the synchrotron radiation and inverse Compton 
scattering  of the pairs with the outer gap accelerator model 
can naturally explain the complex 
 morphology change of the observed pulse profiles. 

We see a small peak feature at the leading phase of 
 Peak~1 of the pulse 
profile of 0.75-10~MeV bands (middle column in Figure~\ref{pulsh}). 
 We will discuss 
this feature in section~\ref{modelp} together with the dependency 
of the results  on the model parameters. In fact, this small peak consists of 
the emissions below the null surface. 

For the polarization characteristics,
Takata et al. (2007) discussed  the predicted characteristics 
 in optical bands  (left column 
in Figure~\ref{pulsl}) 
are in general consistent with the observed features (Kanbach et al. 2005) 
 such as a large position angle swing of the electric vector of the radiation 
at each peak and 
 $\sim 10\%$ of the degree of the polarization  between two peaks. 
In the present study, we find that  
  a similar pattern of the polarization position 
angle (middle panels of Figures~\ref{pulsl} and~\ref{pulsh}) is predicted  
in  the higher energy bands where  the synchrotron process dominates. 
But,  we find that 
the degree of the polarization 
 in the Bridge emissions depends on the photon energy. 
As Figure~\ref{pulsl} shows, the polarization degree
 in the Bridge emissions decreases from $~10\%$ of the optical wavelengths 
with increase of the photon energy. 
This is because the tertiary pairs, 
which  have a small Lorentz factor, contribute 
to the bridge emissions with about $10$~\% of the degree of the polarization  
in the optical wavelengths and contribute with a smaller emissivity 
 in the higher energy region.  In 10-100~MeV bands
 (right column in Figure~\ref{pulsh}), 
where the inverse Compton scattering  dominates,  
 the degree of the polarization  is about ~10\% is predicted.  
By comparing the calculated  position angles 
 of the Bridge emissions below ~1MeV and 
 above 10~MeV,  we find that the present model
 predicts that the observed radiations in 10-100~MeV bands 
are  polarized to the direction  orthogonal to the polarization plane 
of the radiations below 1~MeV. 

 Above 100~MeV, the attenuated curvature photons from the outer gap 
explain the observation as we discussed in section~\ref{dynamics}.
Figure~\ref{curvpuls} summarizes the polarization characteristic of the 
curvature radiation, where we take into
 account the acceleration of the particle 
motion due to the  
curvature of the field lines and of the co-rotating motion for calculating 
the polarization characteristic. We assume  80~\% of the polarization 
degree for each radiation.  The gyration motion can be ignorable in the 
outer gap accelerator because the pitch angle of the primary pairs 
becomes  very small $\sin\theta_p\sim 10^{-7}$. 
We can see from Figure~\ref{curvpuls} that 
the intrinsic level of the degree of the polarization
 at the radiating point is preserved in the Bridge phase 
for the curvature radiation and a small depolarization appears 
around the peaks.
 This less depolarization is because the curvature radiation does not 
related with  the gyration motion of the  pairs. 
  This characteristics of the polarization degree of the curvature radiation 
 were also predicted by  the caustic model in  Dyks et al (2004).
The present model predicts that the radiation above 100~MeV 
of the curvature radiation 
 is highly polarized  more than the radiation below 100~MeV of the synchrotron 
and inverse-Compton process.

\subsubsection{Phase-resolved spectra}
\label{phase}
Figure~\ref{phase50} compares  the predicted and observed phase-resolved 
spectra of the emissions of the Peak~1 (left), Bridge (middle) and 
 Peak~2 (right) phases for the Crab pulsar. 
 The dashed-line and the dotted-line are, 
respectively, the spectra of the synchrotron emissions of the secondary pairs 
beyond and below the null surface. 
 The dashed-dotted line represents the spectrum 
the inverse Compton scattering and includes the emission 
both beyond and below null surface, and the dashed-dotted-dotted line 
represents the appearance curvature spectrum of the primary particles
 in the gap .
The solid-line represents the spectrum of the total emissions.  . 
 To compare the results with the data (Kuiper et al. 2001), 
we normalize the calculated spectrum of Peak~1 to data, but 
we did not normalize individually on the pulse phase.

\begin{figure}
\plotone{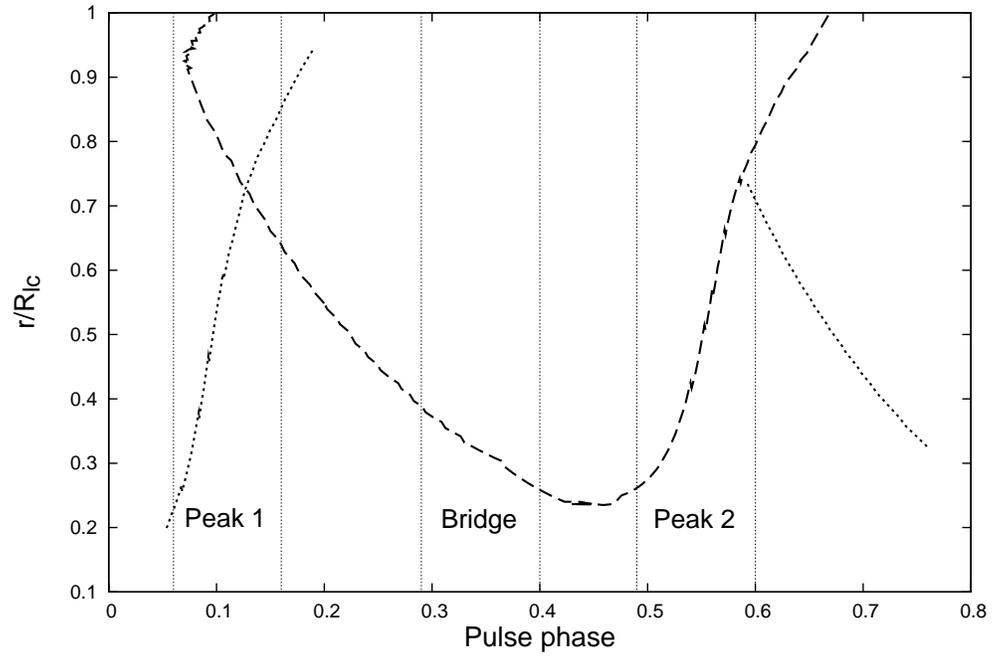}
\caption{Variation of radial distance with pulse phase for the
inclination angle $\alpha=50^{\circ}$ and the 
viewing angle $\xi=100^{\circ}$. The dashed-line and the dotted-line 
represent  the emission regions beyond and below the null surface,
respectively.
 \label{emire}}
\end{figure}

\begin{figure}
 \includegraphics[height=15cm,width=15cm]{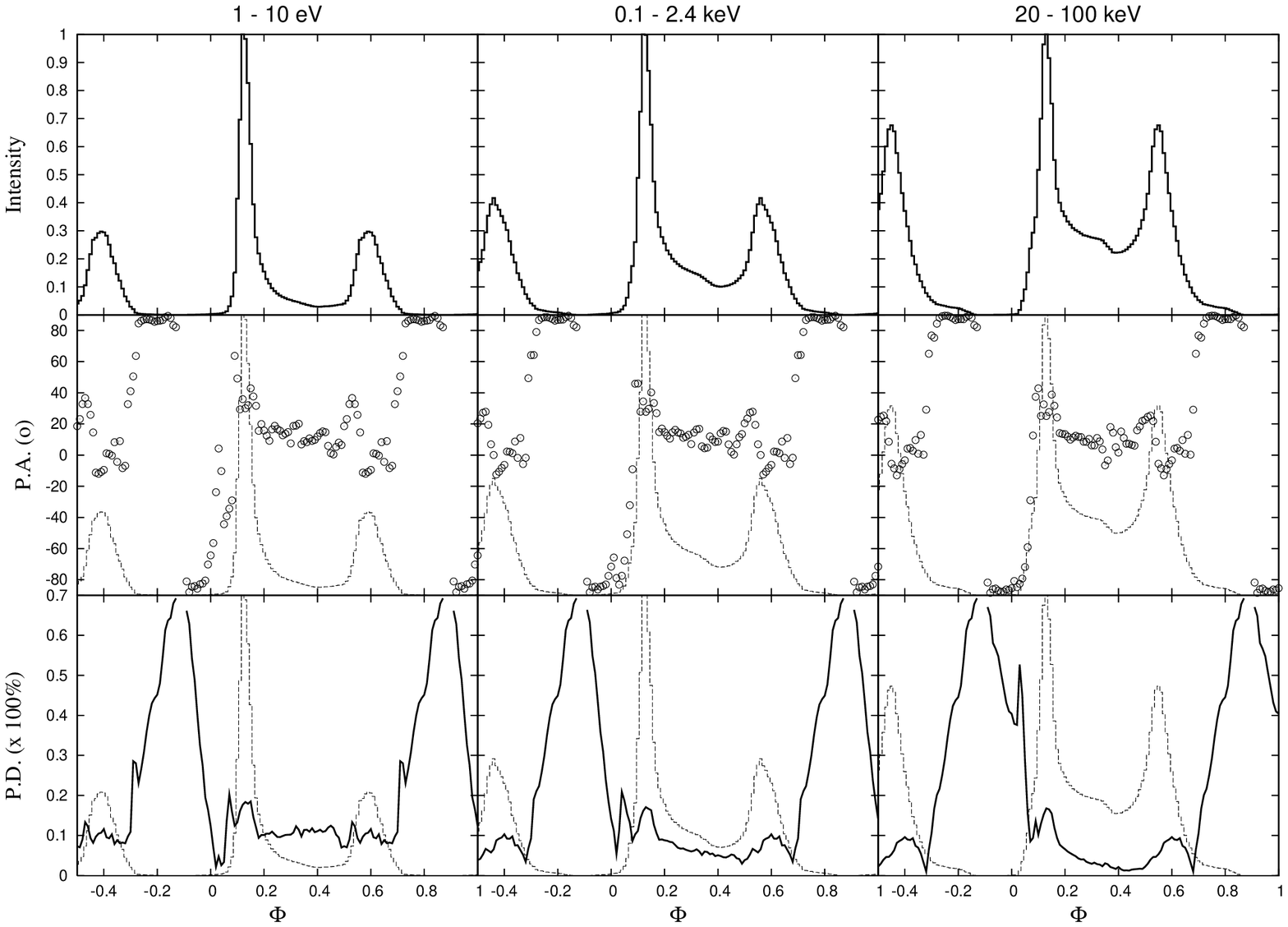}
\caption{Variation of the intensity (tot), the position angle (middle dots)
and the degree
of the polarization (bottom solid-line) 
 as a function of the rotational phase. For the reference, the light
curve is overplotted at middle and bottom panels (thin
dashed-lines). 
To compare with the observe pulse profiles of Kuiper et al. (2001), 
the results were obtained by integrating 
the photons within the energy interval $1-10$~eV (left column), 
$0.1-2.4$~keV (middle column) and  $20-100$~keV (right column).
  The model parameters are
$\alpha=50^{\circ}$, $\xi=100^{\circ}$, $r_{in}=0.67r_n$ and $a_f=1$.
 \label{pulsl}}
\end{figure}

\begin{figure}
 \includegraphics[height=15cm,width=15cm]{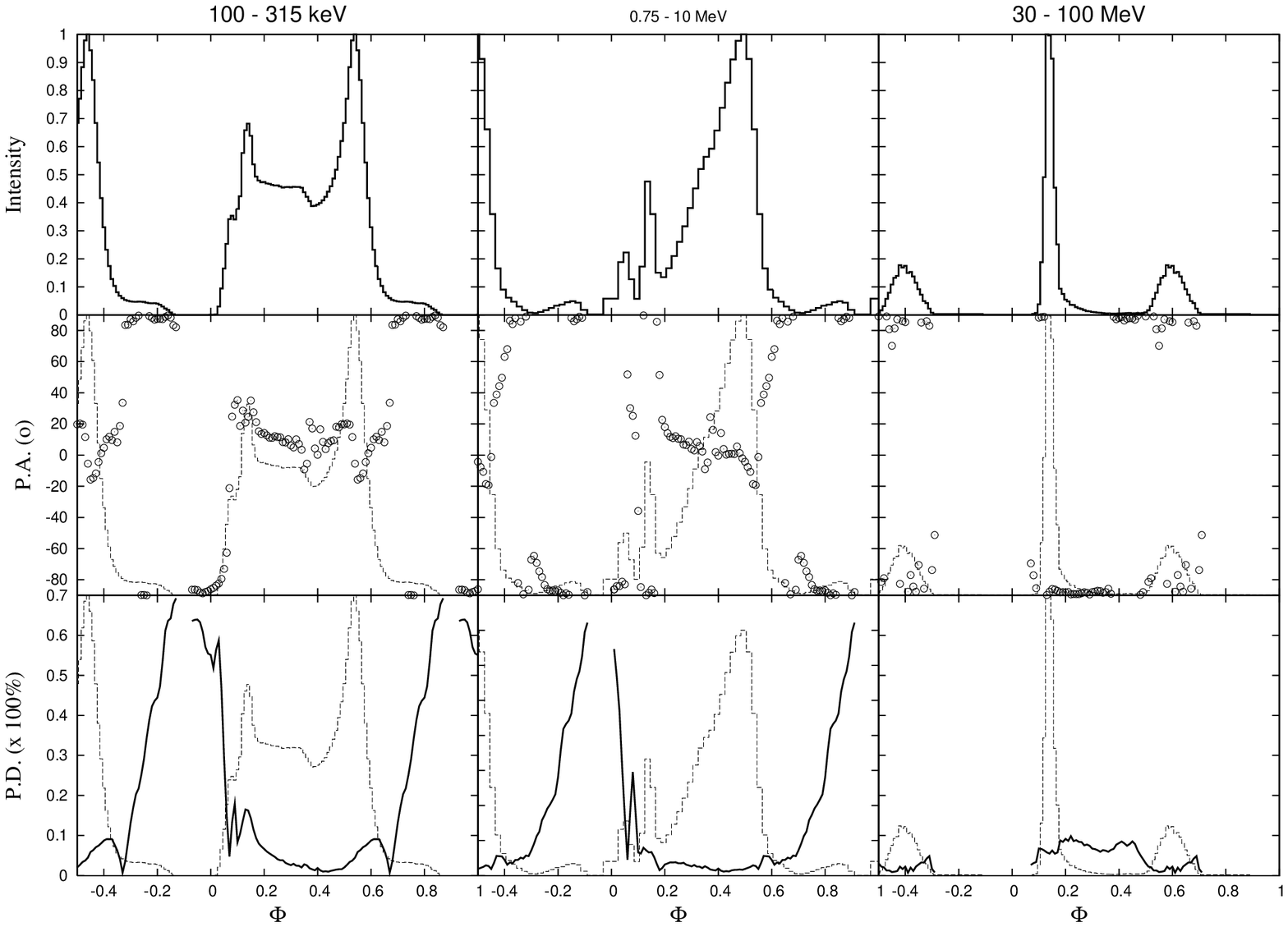}
\caption{Same with the Figure~\ref{pulsl}, but the results are for 
the interval of photons energy, 
 100-315~keV (left), 0.75-10~MeV (middle) and 30-100~MeV (right)\label{pulsh}}
\end{figure}

\begin{figure}
\epsscale{.80}
\plotone{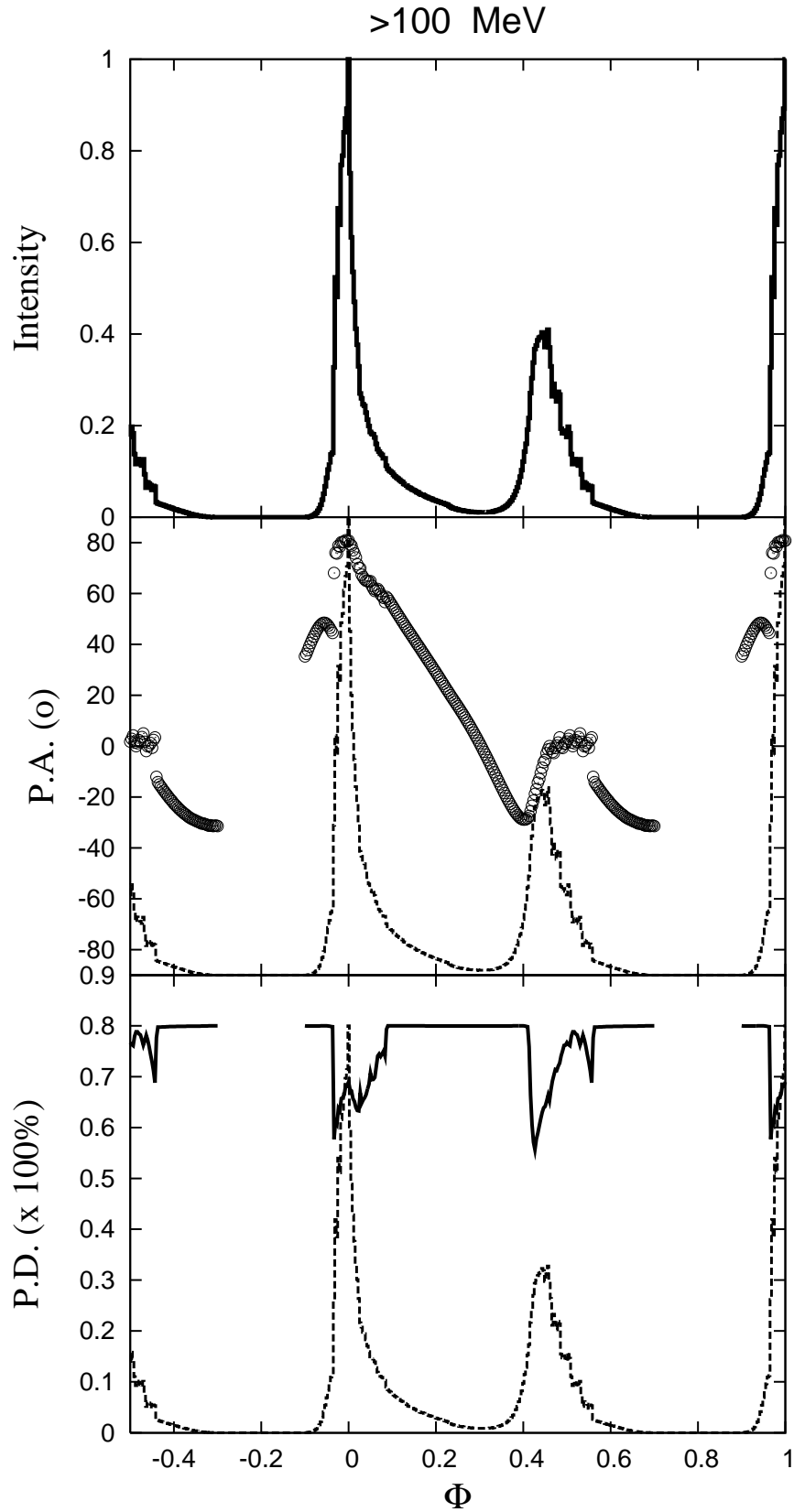}
\caption{The polarization characteristic of the curvature 
radiation in the outer gap. The 80~\% of the polarization degree 
for each radiation was assumed as the intrinsic level.  The results are for 
 $\alpha=50^{\circ}$, $\xi=100^{\circ}$, $r_{in}=0.67r_n$. 
\label{curvpuls}}
\end{figure}

\begin{figure}
 \includegraphics[height=7cm,width=15cm]{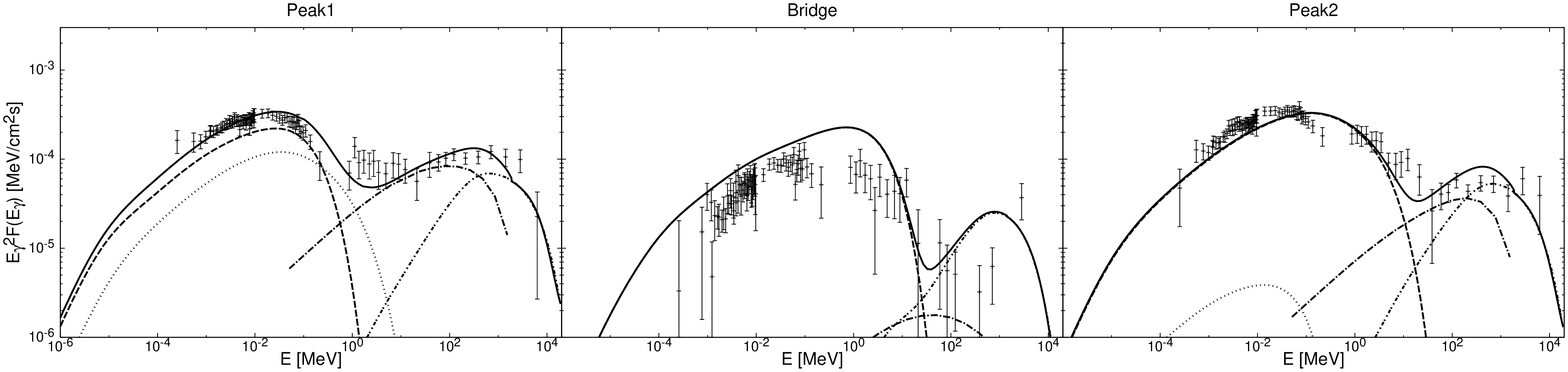}
\caption{Phase-resolved spectra for Peak~1 (left), Bridge (middle) and 
Peak~2 (right) for the Crab pulsar. The dashed-line and the
dotted-line show the spectra of the synchrotron emissions of the pairs
beyond and below the null surface, respectively, the dashed-dotted-line
show the spectrum of the inverse Compton scattering, 
and the dashed-dotted-dotted-line show the attenuated curvature 
spectrum in the outer gap.  The solid-line
represents the spectrum of the total emissions. 
The model parameters are
$\alpha=50^{\circ}$, $\xi=100^{\circ}$, $r_{in}=0.67r_n$ and $a_f=1$. The 
observation data are taken from Kuiper et al. (2001).
 \label{phase50}}
\end{figure}

From the left panel in Figure~\ref{phase50},
 we see  that  the synchrotron emissions below (dotted-line)
and beyond (dashed-line)  the null charge surface 
contribute to the total emissions of Peak~1 and Peak~2. 
Although the flux of the emissions below the  null surface is a smaller than 
that of the emissions beyond the null surface,  
we find from the figure that the emissions below null surface push up  
the total  flux between 1~MeV and 10~MeV bands for Peak~1 
so that the calculated phase-resolved spectrum is in general 
consistent with the data over the optical to $\gamma$-ray bands.
  If we consider only the emissions beyond the null surface 
likewise the traditional outer gap model, we obtain  
 calculated total  flux of  Peak~1 that is significantly 
 smaller than the observed flux 
between 1~MeV and 10~MeV bands. This is because only emissions
 around the light cylinder contribute to  Peak~1 as the dashed 
line in Figure~\ref{emire} shows. 
Because the typical energy of the synchrotron emissions is below
 1~MeV with $B\sim 10^{6}$~Gauss of the magnetic field around 
the light cylinder, its flux decreases exponentially above 1~MeV 
as dashed-line in the left panel of Figure~\ref{phase50} shows. 
On the other hands, we see that 
the synchrotron  emissions below the null charge surface
 contribute to the emission above 1~MeV. This is  because 
the emission region  below the null surface for  Peak~1
 spreads from near the stellar surface to near the 
light cylinder and because the synchrotron emissions with 
 higher magnetic field near the stellar surface contribute to the spectrum.

For Peak~2 (right panel in Figure~\ref{phase50}), we find that 
the observed emissions below 10~MeV 
are explained by only the synchrotron emissions  beyond the null surface, 
unlike with the Peak 1 emissions. For Peak~2 phase, 
the radiating points from near the stellar surface to near the light
cylinder  contribute to the spectrum 
 as the emissions  beyond the null surface as Figure~\ref{emire} shows. 
Therefore, the spectrum of the 
synchrotron photons emitted beyond the null charge surface extends above 
1~MeV, as dashed line in right panel of  Figure~\ref{phase50} shows.
The emission region  below the null surface locates
 only near the light cylinder so that the emissions 
 (dotted-line in the right column of Figure~\ref{phase50}) is negligible
 compared with the emissions beyond the null surface.

In  the Bridge phase, one can see in Figures~\ref{emire} and~\ref{phase50}
 that the photons emitted only beyond the null surface 
are  measured by the observer with the viewing angle $\xi=100^{\circ}$.  
This is due to the  geometry of the dipole magnetic field, and 
 this feature does not depend on the position 
of the location of the inner boundary, providing that 
 we consider only the  outward radiations.
 Namely, the outward emissions  contribute to only  
the Peak~1,  Peak~2  and off-pulse emissions. 
 Therefore, the present model predicts that the Peak~1 and Peak~2 emissions 
consist of the two components (i.e. emissions below and beyond null
surface), while the Bridge emissions consist of only one component
(i.e. emissions beyond null surface).  
This prediction may be consistent with  
 the observational spectral  study done by Massaro et al. (2006), 
 who fitted  the phase-resolved spectra  below 10~MeV 
with specific two functions for  Peak~1 and Pea~2, while one function
 for Bridge. 

When we normalize the model phase-resolved spectra to the observation 
of Peak~1, we find that the calculated flux of the synchrotron emission 
in  Bridge becomes large compared 
with the observation as middle panel in Figure~\ref{phase50} shows. 
We consider that this problem will be related with 
 the three-dimensional structure of the emission region (section~\ref{modelp}).

Between  10~MeV and 100~MeV bands, the emissions from the inverse-Compton 
scattering explain the observation. 
 The present model predicts
 the flux ratio of 1~MeV emissions and 100~MeV emissions in Bridge phase
 is larger than those in Peak~1 and Peak~2.
 This feature is  consistent with the observations.
  With the present synchrotron emission and the inverse-Compton scattering 
model, the flux ratio of 1~MeV emissions and the 100~MeV emissions is 
proportional to square of the strength of the magnetic field. 
 The average magnetic field 
in the  emission region  for  Bridge is larger than  
those for  Peak~1 and Peak~2, because the emission 
region of Bridge  locates nearer the stellar surface than  
those of the Peak~1 and Peak~2 (Figure~\ref{emire}). As a result, 
the predicted flux ratio is larger in the Bridge phase  
than the Peak~1 and Peak~2 phases. 

From Figure~\ref{phase50}, above 500~MeV bands, 
the attenuated curvature radiation from the outer gap 
(dashed-dotted-dotted-line) explains observation.

\subsection{PSR B0540-69}
\label{like}
PSR B0540-69 is sometimes referred to as a twin of the Crab pulsar. 
Its rotational period, surface magnetic field, and spin down luminosity
are similar to those of the Crab pulsar. 
Both pulsar have  similar spectral features in X-ray bands. 
  The observed flux ratio ($<0.01$) of PSR B0540-69 
to the Crab pulsar will represent  ratio of the distances to the pulsars, 
 which are $d\sim49$~kpc
 for PSR B0540-69 and $d\sim2$~kpc for the Crab pulsar. 
Furthermore, the radio giant emissions, which may be associate 
with the high-energy emissions, have been observed from both pulsars 
(Johnston et al. 2004). While they have theses similarities, 
 the observed pulse profiles 
 are quite different to each other. 
The pulse profile of PSR B0540-69 consists of a single broad peak and 
top of the pulse shape becomes relatively flat, whereas 
the pulse profile of the Crab pulsar 
shows a sharp double-peak structure. 
 de Plaa et al (2003) suggested that the observed
 pulse profile of PSR B0540-69 is result of two pulses separated 
~0.2 in phase. These difference will be useful to examine the model capability.

The magnetospheric $X$-ray photon number 
density will be proportional to $n_x\propto L_{gap}/R_{lc}^2$, 
where $L_{gap}\sim f(R_{lc}/2)^3L_{sd}$ and 
$L_{sd}\sim 3.85\times 10^{31}P^{-4}B^2$~erg/s is the spin down energy. 
The ratio of the value of the mean free path, $\lambda\propto 1/n_x$,  
of a primary curvature photon in the magnetospheres of   
PSR B0540-69 to the Crab pulsar 
 is estimated as $\sim 2$. Because the thickness of the emission 
region of the secondary pairs is expected to be related with 
the mean free path of the pair-creation, 
PSR B0540-69 will have a thicker emission 
regions of the secondary pairs than the Crab pulsar. 
   
\begin{figure}
\epsscale{1.20}
\plottwo{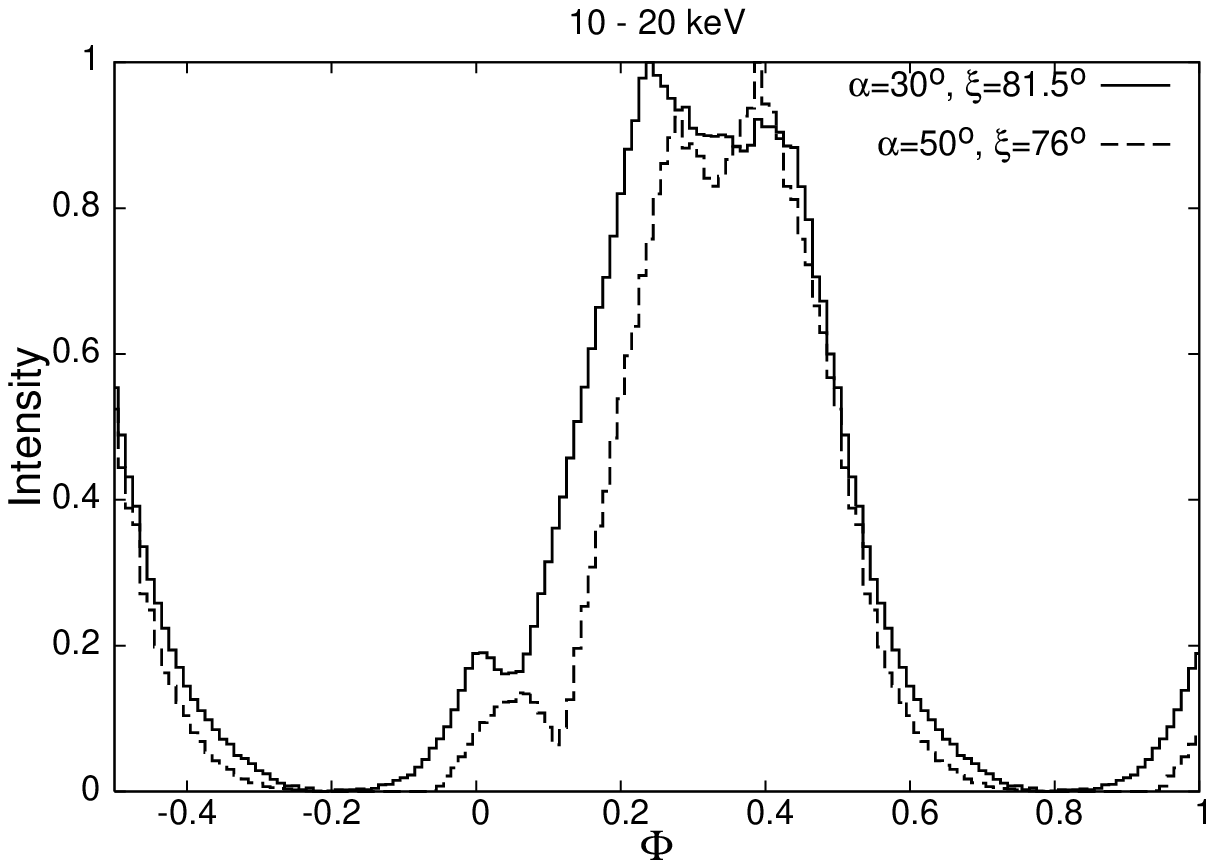}{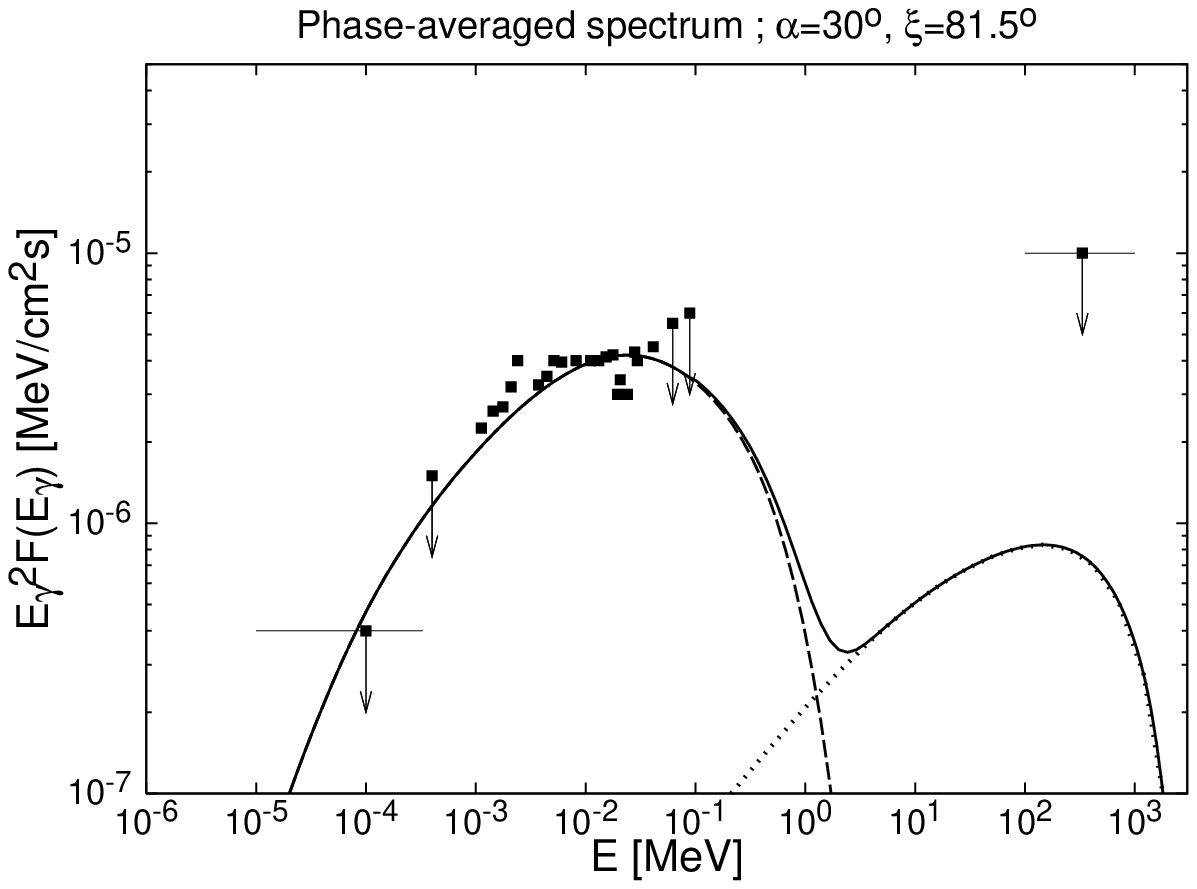}
\caption{Left; The pulse profile in 10-20~keV bands for PSR B0540-69. 
The solid-line is result for  $\alpha=30^{\circ}$ and  $\xi=98.5^{\circ}$ 
(and $\xi=81.5^{\circ}$), 
and the dashed-line is for   $\alpha=50^{\circ}$ and  $\xi=104^{\circ}$ 
(and $\xi=76^{\circ}$). 
Right; The phase-averaged spectrum of PSR B0540-69 for $\alpha=30^{\circ}$ and 
$\xi=98.5^{\circ}$ (and $\xi=81.5^{\circ}$). The solid-line represents 
 the total emissions of the synchrotron radiation (dotted-line) and 
the inverse-Compton scattering (dotted-line). The observation data are take 
from de~Plaa et al (2003). 
\label{B05}}
\end{figure}

Zhang \& Cheng (2000) set the gap thickness of PSR B0540-69 as $\lambda\sim
0.21R_{lc}$, and they chose the inclination  angle $\alpha\sim 50^{\circ}$ 
 and the viewing angle  $\xi=76^{\circ}$. 
 The dashed-line in left panel of Figure~\ref{B05}  shows the pulse profile 
in 10-20~keV bands for the inclination angle $\alpha\sim 50^{\circ}$ 
and for the viewing angle 
angle $\xi=76^{\circ}$ and $\xi=104^{\circ}$. The present model 
 produces a broad single peak in the pulse profile as Zhang \& Cheng (2000) 
showed.   In this viewing angle, however, 
we see that the flat feature of the top of 
the pulse profile appear with a shorter width ($\sim 0.1$ in rotational phase) 
 than the observational value $\sim 0.2$. 
 A  viewing angle more far from 
$90^{\circ}$ produces a more  narrow pulse.
On the contrary,  if we adopt a closer viewing angle to $90^{\circ}$, 
the produced pulse profile has two distinct 
sharp peaks.

With the dipole magnetic field configuration, 
a more consistent results with the observation is obtained 
with a smaller inclination angle than $\alpha=50^{\circ}$. 
The solid-line of the left panel in Figure~\ref{B05}
 shows the pulse profile of 10-20 keV bands 
with the inclination angle $\alpha=30^{\circ}$ and the viewing angle 
$\xi=98.5^{\circ}$ (and $\xi=81.5^{\circ}$). We see that a wider pulse width 
 is produced and the shape of the top of the pulse profile is 
more similar with the observed profile (see figure~2 of de~Plaa et al. 2003).
Therefore, the present model predicts that PSR B0540-69 has 
a thicker emission region (about factor two) than   the Crab pulsar 
and a smaller inclination angle than $\alpha=50^{\circ}$.

\begin{figure}
\epsscale{.80}
 \includegraphics[height=15cm,width=15cm]{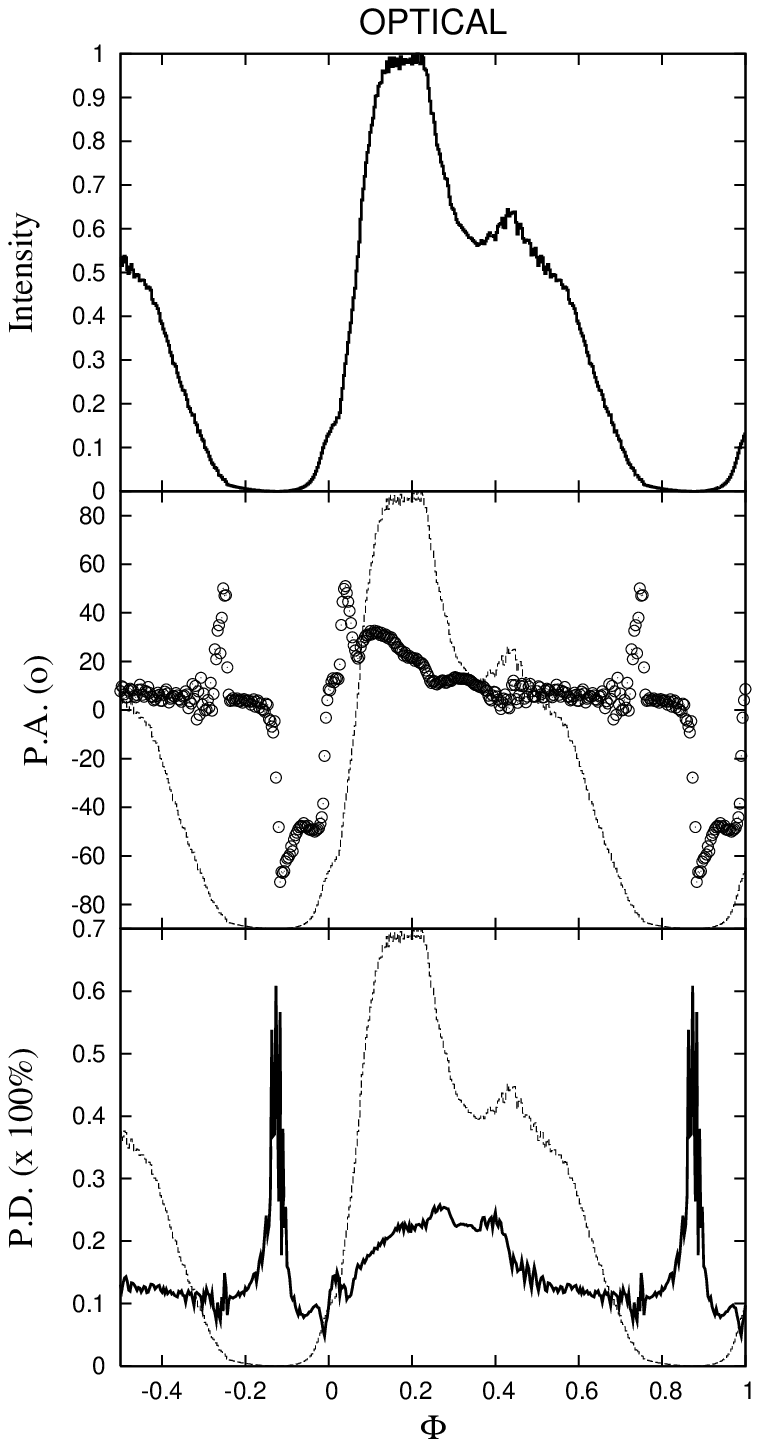}
\caption{Pulse profile in optical bads of PSR B0540-69. The result is for 
 $\alpha=30^{\circ}$ and  $\xi=98.5^{\circ}$. The top, middle
 and bottom panels 
show the intensity, the position angle of the polarization and the degree of 
the polarization, respectively.
 \label{B05op}}
\end{figure}

\begin{figure}
\plottwo{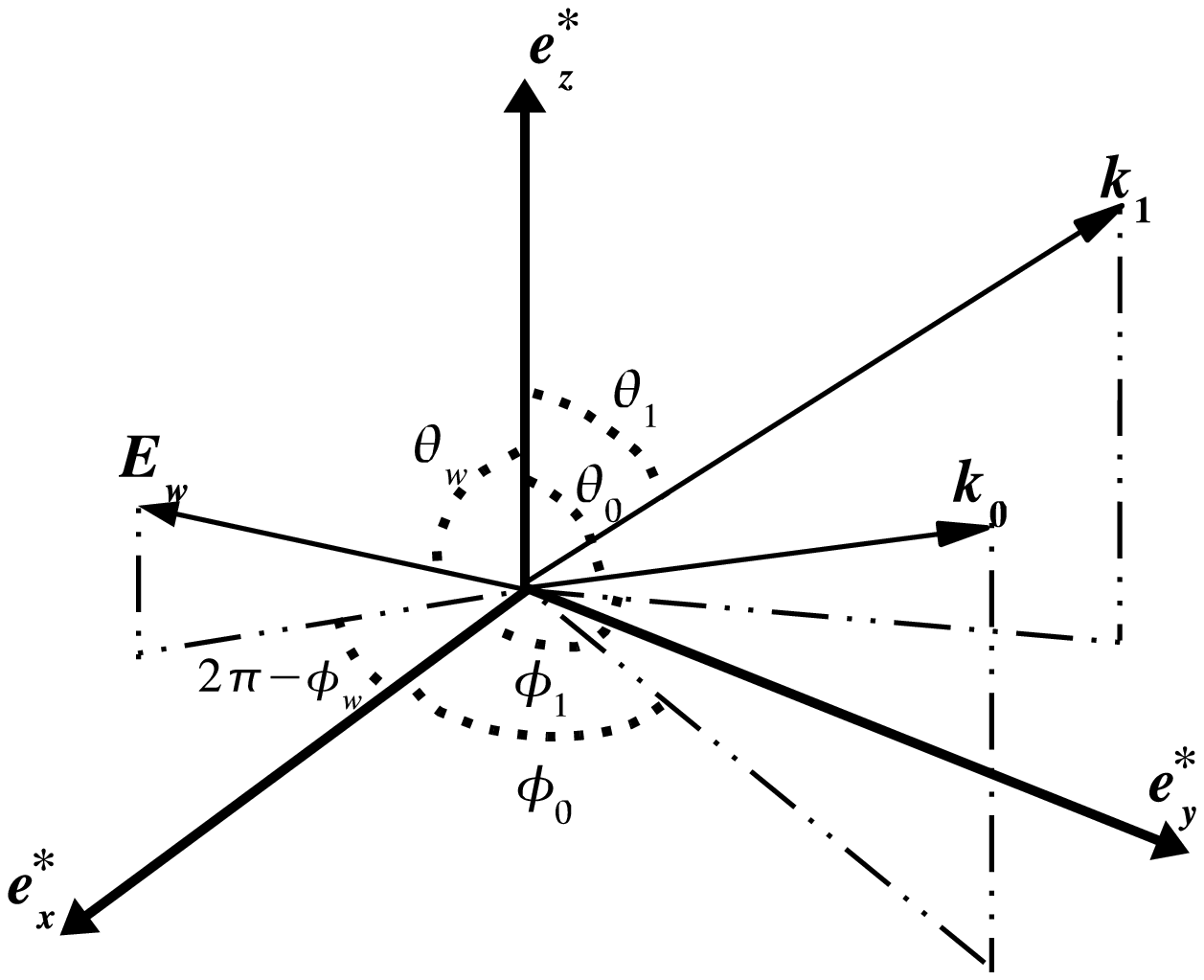}{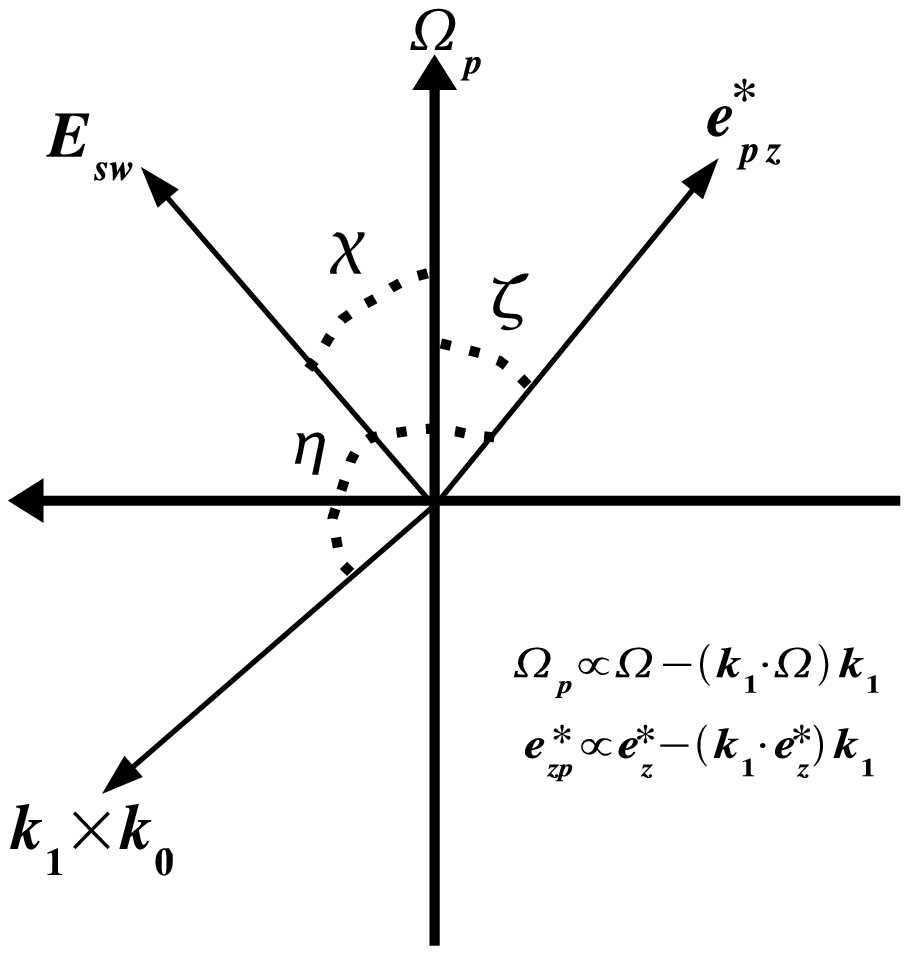}
\caption{Left: Coordinate system used to describe scattering in the electron
rest frame. The polar axis,  $\mbox{\boldmath$e$}^{\ast}_z$, is chosen
to coincide with direction of the electron motion in the observer
frame. The definition of $\mbox{\boldmath$e$}^{\ast}_x$ and 
$\mbox{\boldmath$e$}^{\ast}_z$ are seen in equation~(\ref{unitxy}). 
The vectors $\mbox{\boldmath$k$}_0$ and
$\mbox{\boldmath$k$}_1$
 are the directions of the background and the scattered radiations, 
respectively.  If the background radiation is polarized,  
the vector $\mbox{\boldmath$E$}_w$ is used to 
represent the polarization plane of it.  
Right: Coordinate system  on the sky. $\Omega_{p}$ and 
$\mbox{\boldmath$e$}^{\ast}_{zp}$  is, respectively, 
 the directions of the rotational axis and  of the particle motion 
projected on the sky. The position angle of the 
polarization plane ($\mbox{\boldmath$E$}_{sw}$) is measured anticlockwise from 
the direction of the rotational axis on the sky. 
\label{coord}}
\end{figure}

The expected phase-averaged spectrum in optical to $\gamma$-ray bands 
is displayed in right panel of Figure~\ref{B05} and 
is  consistent with the observations except for the UV data. 
Figure~\ref{B05op} shows the calculated pulse profile in the optical bands. 
We see that the pulse width of the optical bands 
becomes wider than that of the  X-ray bands and the narrower first peak 
becomes more stronger than the narrower second peak.  
These features are also consistent with the data. 
 We also display the predicted  polarization characteristics 
in the optical bands for $\xi=98.5^{\circ}$ in Figure~\ref{B05op}. 
 For the viewing angle $\xi=81.5^{\circ}$, which is 
mutually symmetric with respect to the rotational equator with 
 $\xi=98.5^{\circ}$, the expected pulse
profile and the degree of the polarization are identical with those in
Figure~\ref{B05op}, but the position angle of the polarization is
mirror symmetry of that in Figure~\ref{B05op} with respect to the 
equator $0^{\circ}$.

\section{SUMMARY AND DISCUSSION}
\label{modelp}
In summary, we studied non-thermal emission process of the Crab pulsar with 
the two-dimensional electrodynamical and the three-dimensional studies. 
We  calculated the synchrotron and the curvature radiation process of 
the primary particles in the gap using the electrodynamical study 
and showed a result which produces a consistent gamma-ray spectrum with EGRET 
observation.  Base on the two-dimensional electrodynamical study,
we conducted the three-dimensional model to calculate 
 the synchrotron and the inverse
 Compton scattering 
 of the secondary pairs. We computed the 
 pulse profiles,  
spectra and  polarization characteristics in  optical to
 $\gamma$-ray bands to compare the data of the Crab pulsar and PSR B0540-69. 
 In the three-dimensional study,  we demonstrated that 
the synchrotron radiation and the inverse Compton scattering with 
the outer gap geometry  naturally explain the morphology 
change of the observed pulse 
profiles for the Crab pulsar. 
We also showed that the observed phase-resolved 
spectra of the Crab pulsar 
are explained by the synchrotron radiation 
and the inverse Compton scattering 
of the pairs below and beyond the null surface.
The predicted polarization characteristics of the Crab pulsar  were  
in general consistent with the data in the optical wavelengths. 
We also found that  a broad single pulse profiles and phase-averaged spectrum
 of PSR B0540-69 are explained by the present outer gap 
 model with a thicker emission 
region and a smaller inclination angle than the Crab pulsar.

We have shown that the present outer gap model can explain a lot of 
the observational properties.   But, a further discussion 
for the emission of soft $\gamma$-ray  regions will be required, 
 because  we found that the 
calculated pulse profiles of soft $\gamma$-ray regions, say 0.75-10~MeV, 
easily has  a triple-peak structure in a single period,   
 as the calculated pulse profile 
of 0.75-10~MeV in Figure~\ref{pulsh} indicates. 
In this section, therefore, we discuss the pulse profile of the 
0.75-10~MeV bands together with 
 the dependency of the results on the model parameters. 

As Figure~\ref{emire} shows, 
the both emissions beyond and below the null surface contribute to Peak~1. 
In this case, the small leading peak, which leads the mail first peak, 
 easily appears 
in the pulse profile of 0.75-10~MeV, 
because  the synchrotron spectrum below the null surface is harder 
than that below the null surface. 
Below  MeV bands, the intensity of the  small peak  
is much weaker than that of the main first peak,  and therefore it can not be 
identified in the pulse profile.  However, 
the emissions beyond null surface, which are major contribution of 
the Peak~1 emissions,  are soft so that the intensity 
decreases exponentially above 1~MeV as the calculated phase-resolved 
spectrum (the dashed  line in left panel of the Figure~\ref{phase50}) shows. 
On the other hands, the spectrum of the  emissions below the null surface 
extends above 1~MeV because the photons were emitted near the stellar surface. 
As a result, the peak by the synchrotron emissions below the null surface
 becomes conspicuous in  the pulse profile above 1~MeV. 

 With  a viewing angle more close to $90^{\circ}$ than 
the present $\xi=100^{\circ}$,  one can obtain a pulse profile with 
only two peaks in over the energy bands. 
In this case, the leading small peak
 appears the same phase  with and are not discriminated from Peak~1
  even in the pulse profile of 0.75-10~MeV bands.  
In such a viewing angle, however, the phase separation between the main 
two peaks (Peak~1 and Peak~2) 
 is equal to or larger than $0.5$~phase.   
For a viewing angle more far from  $90^{\circ}$, on the other hands,  
the triple-peak structure  appears not only in the pulse 
profile of 0.75-10~MeV bands, but also the pulse profiles 
in lower energy bands. On these ground, 
we consider the viewing angle $\xi\sim100^{\circ}$ is better fit parameter for 
the inclination angle of $\alpha=50^{\circ}$. 

The altitude of the magnetic surface for the lower boundary 
of the pair emission region 
(i.e. the upper boundary of the outer gap) also affects to the 
pulse profile.
When we adopt a more  higher altitude as the position of 
the boundary than the present position 
(i.e. when we adopt a smaller value of  $a_f=\theta_u/\theta_{lc}$ 
than  the present $a_f=1$),  the phase separation between the leading small 
peak and the main first peak  becomes wider.  
As a result, the pulse profiles have three peaks in a single period from 
optical to $\gamma$-ray bands. 
 On the contrary, when we adopt 
 a lower altitude for the position of the boundary 
 with  a fraction angle  $a_f$ larger than unity, 
we obtain the pulse profiles having only two main peaks, but the phase 
separation between two-peaks becomes to be equal to or larger 
than $0.5$~phase. As the lower boundary of the secondary emission region,  
therefore, we conclude that 
  a magnetic surface close to the surface of the conventional 
last-opened field lines, $a\sim 1$, is 
preferred for the lower boundary of the emission region.
We also examined the different inclination angle. For example, 
the inclination angle $\xi=60^{\circ}$, the viewing angle $\xi\sim100^{\circ}$ 
and $a_f=0.99$ reproduced the consistent phase-resolved spectra and the
 pulse profiles with the Crab data. However, we also obtain  
a small leading  peak in the pulse profile of the 0.75-10~MeV.  

Because the phase of the peak in the pulse profile 
depends on the configuration of 
the magnetic field, the present results would indicate
 that the actual magnetic 
field in the pulsar magnetosphere is modified on important level 
from the vacuum dipole by plasma effects (Muslimov \& Harding 2005). 
To examined this issue, 
we need to solve the magnetic field configuration with the current. 

A more quantitative arguments in the three-dimensional structure of 
the magnetosphere will be required to resolve also  issue that 
 the calculated Bridge emissions are  
large relatively to the observed emissions 
in the soft $\gamma$-ray bands, 
when we normalize the calculated flux using  the Peak~1 emissions, 
as Figure~\ref{phase50}  shows. 
This will indicate that the dependence of the 
 trans-field thickness of the emission region 
is described as  a more complex function of the azimuth  
than the present model, in which the emission region was assumed to be 
located between two  magnetic surfaces. In the actual magnetosphere, 
the trans-field thickness 
around the meridional plane, where the Bridge emissions take place, 
 may be small relative to the thickness of the leading and the
trailing sides, which produce the  Peak~1 and Peak~2 emissions.  This will
be expected because 
the null charge surface 
is located nearer the stellar surface around the meridional plane 
 compared with the leading and the trailing side. Therefore, 
the surface $X$-rays are more dense around the null surface 
in the meridional plane 
 so that the mean-free path of the pair-creation and the resultant 
 trans-field thickness of the outer gap and the emission regions are  
shorter   around the
meridional plane than that of   the leading and trailing
sides. Since the three-dimensional structures  
of the acceleration region and the emission region have not been solved up to 
now, the above issues will demand us to perform a more quantitatively 
study of the three-dimensional structure in the subsequent papers.

Finally,  we note that it is important to observationally constrain 
the inclination angle. 
Takata et al. (2007) searched the parameter range to explain the 
polarization characteristics and the pulse profile in the optical wavelengths 
of the Crab pulsar.
In their study,   a wide range of the viewing angle for
 each inclination angle was allowed for the possible parameters  
to explain the optical pulse profile and the polarization. 
In this study, we found that 
 the narrower range  of the viewing angle 
can produce  the  consistent pulse profile and phase-resolved
spectra with data in optical to $\gamma$-ray bands.  Therefore,  
if the inclination angle can be  determined in other ways, the viewing angle 
will be strongly constrained  by the present study.

\acknowledgments
Authors appreciate fruitful discussion with K.S. Cheng, K.Hirotani,  
S.Shibata and R.Taam.  Authors also thank anomarous referee for insightful
comments on the manuscript. This work was supported
by the Theoretical Institute for Advanced Research in Astrophysics
(TIARA) operated under Academia Sinica and the National Science 
Council Excellence Projects program in Taiwan administered
 through grant number NSC 96-2752-M-007-001-PAE.

\appendix

\section{Stokes parameter of the inverse Compton scattering}
\label{appen}
In this appendix, we derive the Stokes parameters given by 
equations (\ref{unpst}) and 
(\ref{pst}), which are respectively  the Stokes parameters of the 
inverse Compton process for the unpolarized and the polarized 
background radiation. 

In this appendix, as polar-axis (z-axis), 
we choose  the direction of the particle motion at local point. 
With equation (\ref{pmotion}), 
the unit vector $\mbox{\boldmath$e$}^{\ast}_z$ along the z-axis is 
described by the  unit vectors (without asterisk) in 
the coordinate system based on the rotation axis as 
\begin{equation}
\mbox{\boldmath$e$}^{\ast}_z=\beta_0\cos\theta_p\mbox{\boldmath$b$}
+\beta_0\sin\theta_p\mbox{\boldmath$b$}_
{\perp}+\beta_{co}\mbox{\boldmath$e$}_{\phi},
\end{equation}
where $\theta_p$ is the pitch angle, 
$\mbox{\boldmath$b$}$  is the unit vector along the magnetic field,
 and $\mbox{\boldmath$b$}_{\perp}=\pm(\cos\delta\phi\mbox{\boldmath$K$}
+\sin\delta\phi\mbox{\boldmath$K$}\times\mbox{\boldmath$b$})$ 
is the unit vector perpendicular to the 
magnetic field line, $\mbox{\boldmath$K$}$ is the unit vector of the curvature 
of the magnetic field, and $\delta\phi$ is the phase of the gyration motion. 
The parameter $\beta_0$ is determined from $|\mbox{\boldmath$e$}^{\ast}_z|=1$.
W  define $x$ and $y$ axes with (Figure~\ref{coord}) 
\begin{equation}
\mbox{\boldmath$e$}^{\ast}_x=
[\beta_0\mbox{\boldmath$K$}-(\beta_cK_{\phi}
/b_r)\mbox{\boldmath$e$}_r]/|[\beta_0\mbox{\boldmath$K$}
-(\beta_cK_{\phi}
/b_r)\mbox{\boldmath$e$}_r]|,~~\mathrm{and}~~ 
\mbox{\boldmath$e$}^{\ast}_y=\mbox{\boldmath$e$}^{\ast}_z\times
\mbox{\boldmath$e$}^{\ast}_x
\label{unitxy}
\end{equation}
The propagating direction of the background 
radiation,   $\mbox{\boldmath$k$}_0$, are represented   by 
a polar angle $\theta_0$ and a azimuthal angle $\phi_0$, that is, 
$\mbox{\boldmath$k$}_0=\sin\theta_0\cos\phi_0\mbox{\boldmath$e$}^{\ast}_x
+\sin\theta_0\sin\phi_0\mbox{\boldmath$e$}^{\ast}_y
+\cos\theta_0\mbox{\boldmath$e$}^{\ast}_z$.
 The propagating direction of the 
scattered radiation,  $\mbox{\boldmath$k$}_1$, is represented by 
coordinates $(\theta_1,~\phi_1)$. 
If a particle moves with the Lorentz factor $\Gamma_e$, 
we relate 
the  quantities of the rest frame of the electron (with prime) 
and of the observer frame (without prim)  as 
\[
\cos\theta'=(\cos\theta-\beta)/(1-\beta\cos\theta),~~\phi'=\phi, 
\]
for the polar and azimuthal angles and  
$\epsilon=\Gamma(1+\beta\cos\theta')\epsilon'$
for the photon energy. 
If the background radiation is polarized, we describe the direction of 
the polarization with $\mbox{\boldmath$E$}_{w}=\sin\theta_w\cos\phi_w
\mbox{\boldmath$e$}^{\ast}_x+\sin\theta_w\sin\phi_w\mbox{\boldmath$e$}^{\ast}_y
+\cos\theta_w\mbox{\boldmath$e$}^{\ast}_z$.

In the calculation, we first compute the Stokes parameters 
of the Compton process in the electron rest frame, 
and then we perform the Lorentz transformation  to the observer frame.
\subsection{Unpolarized background field}
For the Compton scattering with the unpolarized background radiation,
 the Stokes parameters, which are 
 represented in the coordinate basis orthogonal to and 
parallel to the scattering plane, in the electron rest frame
 are  
\begin{equation}
\left. 
\begin{array}{@{\,}ll}
&s^{u'}_0 \\
&p^{u'}_{\perp} \\
&p^{u'}_{||}
\end{array}
\right\}=
\left\{
\begin{array}{@{\,}ll}
\left[\frac{\epsilon'_0}{\epsilon'_1}+\frac{\epsilon'_1}{\epsilon'_0}
-\sin^2w'_s\right]& \\
\sin^2w'_ s& \\
0&
\end{array}
\right.
\label{resups}
\end{equation}
(McMaster, 1961),  where $s^{u'}_0$ corresponds to the emissivity,
 $p^{u'}_{\perp}$ 
and $p^{u'}_{||}$ describe the polarization 
direction orthogonal to and the parallel to the plane of the scattering, 
respectively.   The dimensionless quantities  $\epsilon'_0$ 
and $\epsilon'_{1}$ are, respectively,  the 
energy of the background and the scattered photons in units
 of the electron rest mass energy and are connected by 
$\epsilon'_1=\epsilon'_0/[1+\epsilon'_0(1-\cos w'_s)]$. The scattering 
angle $w'_s$ is defined by 
\begin{equation}
\cos w'_s=\mbox{\boldmath$k$}'_0\cdot\mbox{\boldmath$k$}'_1
=\sin\theta'_0\sin\theta'_1\cos(\phi_0-\phi_1)+\cos\theta'_0\cos\theta'_1.
\end{equation}
For the unpolarized background radiation, we find that 
the scattered radiation is partially polarized 
orthogonal to the scattering plane.

 We transform the back ground frame to the one 
related with the particle motion 
direction (z-axis) projected onto the sky. Using such coordinate system, 
the Stokes parameters  become
\begin{equation}
q'_u
=p^{u'}_{\perp}\cos2\eta'-p^{u'}_{\perp}\sin2\eta'=\sin^2w'_s\cos2\eta',
\label{supq}
\end{equation}
and 
\begin{equation}
u'_u=p^{u'}_{\perp}\sin2\eta'+p^{u'}_{\perp}\cos2\eta'
=\sin^2w'_s\sin2\eta',
\label{supu} 
\end{equation}
where $\eta'$ is angle between  the projected z-axis on the 
sky and the direction orthogonal to the scattering plane, and  
its cosine is given by $\cos\eta'=-\sin\theta'_0\sin(\phi'_0-\phi'_1)
/\sin w'_s$ (Figure~\ref{coord}).

Writing the unpolarized background radiation in the rest frame
 with  $I^{u'}_b(\epsilon_0,\mbox{\boldmath$k$}'_0)$, 
the Stokes parameters of the total amount of the 
scattered radiation are give by
$i^{u'}(\epsilon'_1,\mbox{\boldmath$k$}'_1)
=(3\sigma_T/16\pi)\int I^{u'}_{b} s^{u'}_0d\Omega'$,
  $Q'_u(\epsilon'_1,\mbox{\boldmath$k$}'_1)
=(3\sigma_T/16\pi)\int I^{u'}_{b} q'_u d\Omega'$, 
and $U'_u(\epsilon'_1,\mbox{\boldmath$k$}'_1)
=(3\sigma_T/16\pi)\int I^{u'}_b u'_ud\Omega'$, respectively. If we take 
the  Thomson limit, $\epsilon'_0=\epsilon'_1$, we arrive
 the equations (2), (4a) and (4b) of Begelman and Sikora (1987).

By performing the Lorentz transformation 
from the electron rest frame to 
observer inertial frame, and by transforming the background coordinate to 
one related the projected rotation axis on the sky, the Stokes parameters 
become 
\begin{eqnarray}
i(\mbox{\boldmath$k$}_1,\epsilon_1)&
=&D^3_2i'^3(\mbox{\boldmath$k$}'_1,\epsilon_1)\,
 \nonumber \\
q(\mbox{\boldmath$k$}_1,\epsilon_1)&=&D^3_2
[q'(\mbox{\boldmath$k$}'_1,\epsilon'_1)\cos2\zeta-
u'(\mbox{\boldmath$k$}'_1,\epsilon'_1)\sin2\zeta], \\
u(\mbox{\boldmath$k$}_1,\epsilon_1)&=&D^3_2
[u'(\mbox{\boldmath$k$}'_1,\epsilon'_1)\cos2\zeta+
q'(\mbox{\boldmath$k$}'_1,\epsilon'_1)\sin2\zeta], \nonumber 
\label{sstoke}
\end{eqnarray}
where $D_2=\epsilon_1/\epsilon'_1=\Gamma_e^{-1}(1-\beta\cos\theta_0)^{-1}$ 
is the Doppler factor, and the azimuth angle $\zeta$ is defined by angle 
between the angle between the direction of the rotation axis and of 
the particle motion  projected on the sky.

Finally, integrating $i$, $q$ and $u$ over the particle momentum distribution, 
the Stokes parameters are given by
\begin{equation}
\left.
\begin{array}{@{\,}ll}
&dI_{u}(\mbox{\boldmath$k$}_1,\epsilon_1)/ds \\
&dQ_{u}(\mbox{\boldmath$k$}_1,\epsilon_1)/ds \\
&dU_{u}(\mbox{\boldmath$k$}_1,\epsilon_1)/ds
\end{array}
\right\}=\int d\Gamma_e 
\left[\frac{dn_e}{d\Gamma_e}\right] (1-\beta\cos\theta_0)
\times 
\left\{
\begin{array}{@{\,}ll}
i(\mbox{\boldmath$k$}_1,\epsilon_1),& \\
q(\mbox{\boldmath$k$}_1,\epsilon_1),& \\
u(\mbox{\boldmath$k$}_1,\epsilon_1),&
\end{array}
\right.
\label{lstoke}
\end{equation}
where the extra factor $(1-\beta\cos\theta_0)$ represents relative 
motion of photons and particles along the photon motion.

Now, we suppose the total background  radiation in the observer frame that 
$I_b(\mbox{\boldmath$k$}_0,\epsilon_0)=C_0\epsilon^{-\alpha}_0
\delta(\theta-\theta_0)
\delta(\phi-\phi_0)$. In the rest frame of an electron, 
 the background radiation distributes as 
\begin{equation}
I'_b(\mbox{\boldmath$k$}'_0,\epsilon'_0)=D_1^3I_b(\mbox{\boldmath$k$}_0
,\epsilon_0)=\frac{C_0\epsilon'^{-\alpha}_0}
{\Gamma_e^{1+\alpha}(1+\beta\cos\theta')^{1+\alpha}}\delta(\theta'-\theta'_0)
\delta(\phi'-\phi'_0),
\label{incid}
\end{equation}
where we use $D_1=\epsilon'_0/\epsilon_0=\Gamma_e^{-1}
(1+\beta\cos\theta')^{-1}$, 
and $\delta(\theta-\theta_0)\delta(\phi-\phi_0)=
D_1^{-2}\delta(\theta'-\theta'_0)\delta(\phi'-\phi'_0)$, 
which is because $\delta(\theta-\theta_0)\delta(\phi-\phi_0)d\Omega
=\delta(\theta'-\theta'_0)\delta(\phi'-\phi'_0)d\Omega'$
 and $d\Omega=D_1^2d\Omega'$.

By denoting  the polarization degree of the total background radiation as  
$\Pi_{syn}$, that is, $I^{u'}_{b}=(1-\Pi_{syn})I'_b$, 
finally we arrive equitation (\ref{unpst}), 
\begin{eqnarray}
\left.
\begin{array}{@{\,}ll}
&dI_u(\mbox{\boldmath$k$}_1,\epsilon_1)/ds \\
&dQ_u(\mbox{\boldmath$k$}_1,\epsilon_1)/ds \\
&dU_u(\mbox{\boldmath$k$}_1,\epsilon_1)/ds
\end{array}
\right\}&=&(1-\Pi)\frac{3\sigma_T}{16\pi}C_0\int d\Gamma_e 
\left[\frac{dn_e}{d\Gamma_e}\right]\frac{\epsilon^{-\alpha}_0}
{\Gamma^{4+\alpha}_e(1-\beta\cos\theta_1)^{2}(1+\beta\cos\theta'_0)^{1+\alpha}}
\nonumber \\
&\times& 
\left\{
\begin{array}{@{\,}ll}
s^{u'}_{0}(\mbox{\boldmath$k$}'_1,\epsilon'_1),& \\
q'_u(\mbox{\boldmath$k$}'_1,\epsilon'_1)\cos2\zeta
-u'_u(\mbox{\boldmath$k$}_1,\epsilon_1)\sin2\zeta,& \\
q'_u(\mbox{\boldmath$k$}'_1,\epsilon'_1)\sin2\zeta
+u'_u(\mbox{\boldmath$k$}_1,\epsilon_1)\cos2\zeta.&
\end{array}
\right.
\label{unpstokes}
\end{eqnarray}

\subsection{Polarized background field}
For the polarization background radiation, the Stokes parameters corresponding 
to equation (\ref{resups}) become
\begin{equation}
\left. 
\begin{array}{@{\,}ll}
&s^{p'}_0\\
&p^{p'}_{\perp} \\
&p^{p'}_{||}
\end{array}
\right\}=
\left\{
\begin{array}{@{\,}ll}
\left[\frac{\epsilon'_0}{\epsilon'_1}+\frac{\epsilon'_1}{\epsilon'_0}
-\sin^2\theta'_1\cos^2\lambda'_p\right]& \\
\sin^2\theta'_1-(1+\cos^2\theta'_1)\cos2\lambda'_p & \\
2\cos\theta'_1\sin2\lambda'_p&
\end{array}
\right.
\end{equation}
 where $\lambda'_p$ is  the angle between the polarization 
plane of the background 
radiation and the orthogonal plane to 
the scattering plane and  its cosine is given by 
\begin{eqnarray}
\cos\lambda'_p&=&(\sin w'_s)^{-1}[\sin\theta'_0\cos\theta'_1\sin\theta'_w
\sin(\phi_0-\phi_w)+\cos\theta'_0\sin\theta'_1\sin\theta'_w 
\sin(\phi_w-\phi_1) \nonumber \\
&+&\sin\theta'_0\sin\theta'_1\cos\theta'_w
\sin(\phi_1-\phi_0), 
\end{eqnarray}
where $\theta'_w$ and $\phi'_w$ represent the polar angle 
and the azimuthal angle of the polarization plane of the background
 radiation, respectively.

Transforming the background coordinate to one related with the particle motion,
the components of the Stokes parameters representing the polarization 
are, respectively,  given by  
\begin{equation}
q'_p=[\sin^2w'_1-(1+\cos^2w')\cos2\lambda'_p]\cos2\eta'
-2\cos'w\sin2\lambda'_p\sin2\eta,
\end{equation}
and 
\begin{equation}
u'_p=[\sin^2w'_1-(1+\cos^2w')\cos2\lambda'_p]\sin2\eta'
+2\cos'w\sin2\lambda'_p\cos2\eta.
\end{equation}
After same procedure with the case of the unpolarized background radiation
 (or replacing subscript $u$  and $1-\Pi_{syn}$ in the 
equation~(\ref{unpstokes}) to $p$ and to $\Pi_{syn}$), 
we arrive equation (\ref{pst}).



\clearpage



\clearpage

\clearpage

\clearpage


\clearpage

\clearpage

\clearpage

\clearpage

\clearpage

\clearpage

\clearpage

\clearpage


\end{document}